\documentclass[12pt,journal,onecolumn,draftcls]{IEEEtran}

\usepackage{epsfig}
\usepackage{times}
\usepackage{float}
\usepackage{afterpage}
\usepackage{amsmath}
\usepackage{amstext}
\usepackage{amssymb,bm}
\usepackage{latexsym}
\usepackage{color}
\usepackage{graphicx}
\usepackage{amsmath}
\usepackage{amsthm}
\usepackage{graphicx}
\usepackage[center]{caption}
\usepackage{pstricks}
\usepackage{subfigure}
\usepackage{booktabs}

\newtheorem{theorem}{Theorem}
\newtheorem{remark}{Remark}

\newtheorem{proposition}{Proposition}
\newtheorem{definition}{Definition}

\def \eu{\mathrm{e}}
\def \jj{\mathrm{j}}
\def \st{\mathrm{s.t.}}

\begin{document}

\title{Gaussian Half-Duplex Relay Networks: improved constant gap and connections with the assignment problem}

\author{Martina~Cardone, Daniela~Tuninetti, Raymond~Knopp and Umer~Salim %
\thanks{M. Cardone and R.  Knopp are with the Mobile Communications Department at Eurecom, Biot, 06410, France (e-mail: cardone@eurecom.fr; knopp@eurecom.fr). D. Tuninetti is with the Electrical and Computer Engineering Department of the University of Illinois at Chicago, Chicago, IL 60607 USA (e-mail: danielat@uic.edu). U. Salim is with the Algorithm Design group of Intel Mobile Communications, Sophia Antipolis, 06560, France (e-mail: umer.salim@intel.com).

The work of D.~Tuninetti was partially funded by NSF under award number 0643954;
the contents of this article are solely the responsibility of the author and
do not necessarily represent the official views of the NSF.
Eurecom's research is partially supported by its industrial partners: BMW Group Research \& Technology, IABG, Monaco Telecom, Orange, SAP, SFR, ST Microelectronics, Swisscom and Symantec. The research at Eurecom leading to these results has received funding from the European Union Seventh Framework Programme under grant agreement CONECT $\rm{n}^\circ$ 257616. The research work carried out at Intel by U. Salim has received funding from the EU FP7 grant agreement iJOIN ($\rm{n}^\circ$ 317941). %(FP7/2007-2013)

M.~Cardone would like to thank Paul de  Kerret for insightful discussions.

The results in this paper were submitted in part to the 2013 IEEE Information Theory Workshop (ITW) \cite{USITW2013multirelay}.}
}
\maketitle
% ----------------------------
\begin{abstract}
This paper considers a general Gaussian relay network where a source transmits a message to a destination with the help of $N$ half-duplex relays. 
%The relays work in half-duplex mode, in the sense that they can not transmit and receive at the same time.
It proves that the information theoretic cut-set upper bound to the capacity can be achieved to within $2.021(N+2)$ bits with noisy network coding, thereby reducing the previously known gap.
%holds for generic Gaussian half-duplex relay networks and 
Further improved gap results are presented for more structured networks like diamond networks. 

It is then shown that the generalized Degrees-of-Freedom of a general Gaussian half-duplex relay network is the solution of a linear program, where the coefficients of the linear inequality constraints are proved to be the solution of several linear programs, known in graph theory as the assignment problem, for which efficient numerical algorithms exist. 
%, or weighted bipartite matching problem. 
The optimal schedule, that is, the optimal value of the $2^{N}$ possible transmit-receive configurations/states for the relays, is investigated and known results for diamond networks are extended to general relay networks. It is shown, for the case of $2$ relays, that only $3$ out of the $4$ possible states have strictly positive probability. Extensive experimental results show that, for a general $N$-relay network with $N\leq 8$, the optimal schedule has at most $N+1$ states with strictly positive probability. As an extension of a conjecture presented for diamond networks, it is conjectured that this result holds for any HD relay network and any number of relays.

Finally, a $2$-relay network is studied to determine the channel conditions under which selecting the best relay is not optimal, and to highlight the nature of the rate gain due to multiple relays.
\end{abstract}

% ----------------------------

\section{Introduction}
\label{sec:intro}
Cooperation between nodes in a network has been proposed as a potential and promising technique to enhance the performance of wireless systems in terms of coverage, throughput, network generalized Degrees-of-Freedom (gDoF) and robustness / diversity. This last point is of great importance, especially in military and satellite communications, where redundancy and diversity play a significant role, by insuring a more reliable link between two networks (military communication) and two ground stations (satellite communication), with respect to the point-to-point communication.

The simplest form of collaboration can be modeled as a Relay Channel (RC) \cite{coveElGamal}. The RC is a multi-terminal network where a source conveys information to a destination with the help of one relay. The relay has no own data to send and its only purpose is to assist the source in the transmission. 
%Thanks to the different connections between the nodes, the use of many relays leads to an increase in the robustness and network gDoF. 
Motivated by the undeniable practical importance of the RC, in this paper we analyze a system where the communication between a source and a destination is assisted by multiple relays. In particular, we mainly focus on the enhancement in terms of %diversity and 
gDoF due to the use of multiple Half-Duplex (HD) relays. A relay is said to work in HD mode %, e.g., through time- or frequency-division, 
if at any time / frequency instant it can not simultaneously transmit and receive. The HD modeling assumption is at present more practical than the Full-Duplex (FD) one. 
%, at least with  current technologies
This is so because practical restrictions arise when a node can simultaneously transmit and receive, such as for example how well self-interference can be canceled, making the implementation of FD relays challenging \cite{Duarte,Everett}. 
%FD may be achieved with deployment constraints, for instance, separate transmit and receive antennas with sufficient spacing as well as self-interference cancellation through sophisticated signal processing techniques.

\subsection{Related Work}
The RC model
%, where a source communicates with a destination via the help of a relay node,
was first introduced by van der Meulen \cite{Meulen} in 1971. Despite the significant research efforts, the capacity of the general RC is still unknown. 
%The most important capacity results on the RC are due to Cover and El Gamal~\cite{coveElGamal}. 
In their seminal work~\cite{coveElGamal}, Cover and El Gamal proposed a general outer bound, now known as the {\em max-flow min-cut outer bound} or cut-set for short, and  two achievable schemes: {\em decode-and-forward} (DF) and {\em compress-and-forward} (CF). The cut-set outer bound was shown to be tight for the degraded RC, the reversely degraded RC and the semi-deterministic RC~\cite{coveElGamal}, but it is not tight in general~\cite{AleksicRazaghiYu}.
%In~\cite{coveElGamal}, two achievable schemes were proposed as well, i.e., the {\em decode-and-forward} (DF) and the {\em compress-and-forward} (CF), whose combination is still the best known achievable strategy for the general RC.

%Usually the relay is assumed to operate in FD. 
%%This assumption may not be realistic due to practical restrictions such as the inability to perfectly cancel the self-interference. Thereby, 
%However due to practical restrictions, HD relaying techniques have also been studied. 
Although more study has been conducted for FD relays, there are some important references threating HD ones.
In \cite{Host2002}, the author studied the time-division duplexing RC.
% where, at each time-slot, the relay is in receiving mode for a fraction $\alpha$ of time and in the transmitting mode for the remaining $1-\alpha$ fraction. 
Both an outer bound, based on the cut-set argument, and an inner bound, based on {\em partial decode-and-forward} (PDF) were developed. In \cite{Host2002}, the time instants where the relay switches from listen to transmit and vice versa are assumed fixed, i.e., a priori known by all nodes; 
%In other words, the source and the destination are aware about the activity (either listening or transmitting) of the relay. 
we refer to this mode of operation as {\em deterministic switch}. 
In \cite{kramer-allerton}, it was shown that higher rates can be achieved by considering a {\em random switch} at the relay. In this way the randomness that lies into the switch  %from listen to transmit states
 may be used to transmit (at most one bit per channel use of) further information to the destination. In \cite{kramer-allerton}, it was also shown how the memoryless FD framework incorporates the HD one as a special case, and as such there is no need to develop a separate theory for networks with HD relays. 

The pioneering work of~\cite{coveElGamal} has been extended to networks with multiple relays. In~\cite{ KramerGastparGuptaIT05}, the authors proposed several inner and outer bounds for FD relay networks as a generalization of DF, CF and the cut-set bound; it was shown that DF achieves the ergodic capacity of a wireless Gaussian network with phase fading if phase information is locally available and the relays are close to the source node. 

The exact characterization of the capacity region of a general memoryless network is challenging. Recently it has been advocated that progress can be made towards understanding the capacity by showing that achievable strategies are provably close to (easily computable) outer bounds \cite{AvestimehrThesis}. As an example, 
%Etkin {\em et al.} characterized the capacity of the Gaussian Interference Channel to within 1 bit regardless of the system parameters~\cite{WangTse}.
in \cite{avestimher:netflow}, the authors studied FD %unicast and multicast
Gaussian relay networks with $N+2$ nodes (i.e., $N$ relays, a source and a destination) and showed that the capacity can be achieved to within $\sum_{k=1}^{N+2}5\min\{M_k,N_k\}$ bits with a network generalization of CF named {\em quantize-remap-and-forward} (QMF), where $M_k$ and $N_k$ are the number of transmit and receive antennas, respectively, of node $k$. Interestingly, the gap result remains valid for static and ergodic fading networks where the nodes operate either in FD mode or in HD mode; however \cite{avestimher:netflow} did not account for random switch in the outer bound. 
In \cite{Ozgur2010}, the authors demonstrated that the QMF scheme can be realized with nested lattice codes. %, both in the transmission and in the quantization phases,  still achieves the Gaussian capacity to within a constant gap. Interestingly they used the approach 
Moreover they showed that for single antenna HD networks with $N$ relays, by following the approach of \cite{kramer-allerton}, i.e., by also accounting for random switch in the outer bound, the gap is $5N$ bits.
Recently, for single antenna networks with $N$ FD relays, the $5(N+2)$~bits~gap of \cite{avestimher:netflow} was reduced to $2 \times 0.63(N+2)$~bits (where the factor $2$ accounts for complex-valued inputs) thanks to a novel ingenious generalization of CF named {\em noisy network coding} (NNC) \cite{nncLim}.
% in the derivation of the cut-set bound, by taking into account a random switch at the relays. For single antenna systems, Lim {\em et al.} in~\cite{nncLim} recently showed that this . Both QMF and NNC are network extensions of CF.
 
The gap characterization of~\cite{nncLim} is valid for a general Gaussian network with FD relays; the gap grows linearly with the number of nodes in the network, which could be a too coarse capacity characterization for networks with a large number of nodes. Smaller gaps can be obtained for more structured networks. For example, a {\em diamond network}~\cite{ScheinGallagerISIT200} consists of a source, a destination and $N$ relays where the source and the destination can not communicate directly and the relays can not communicate among themselves. In other words, a general Gaussian relay network with $N$ relays is characterized by $(N+2)(N+1)$ generic channel links, while a diamond network has only $2N$ non-zero channel links. In~\cite{ScheinGallagerISIT200} the case of $N=2$ relays was studied and an achievable region based on time sharing between DF and {\em  amplify-and-forward} (AF) was proposed.
The capacity of a general FD diamond network is known to within $2\log(N+1)$~bits~\cite{arXiv:1207.5660}. 
If, in addition, the FD diamond network is symmetric, that is, all source-relay links are equal and all relay-destination links are equal, the gap is less than 2~bits for any $N$~\cite{NiesenDiggavi}. 

HD diamond networks have been studied as well. %, albeit mainly with deterministic switch at the relays.
In a HD diamond network with $N$ relays, there are $2^{N}$ possible combinations of listening / transmitting states, since each relay, at a given time instant, can either transmit or receive. For the case of $N=2$ relays, \cite{Bagheri2009} showed that out of $2^{N}=4$ possible states only $N+1=3$ states suffice to achieve the cut-set upper bound to within less than 4~bits; we refer to the states with strictly positive probability as {\em active states}.
The achievable scheme of \cite{Bagheri2009} is a clever extension of the two-hop DF strategy of~\cite{XueSandhuIT07}. In \cite{Bagheri2009} a closed-form expression for the aforementioned active states, by assuming no power control and deterministic switch, was derived by solving the dual linear program (LP) associated with the LP derived from the cut-set upper bound.
%and tight outer bounds based on the dual of the linear program (LP) associated with the classical cut-set bound. Extensions of these ideas to more than two relays appear difficult due to the combinatorial structure of the problem.
The work in~\cite{Fragouli2012}
%, the authors adapted the cut-set bound derived in \cite{Ozgur2010} to the HD diamond network. They 
considered a general diamond network with $N=2$ relays and an `antisymmetric' diamond network with $N=3$ relays 
%{\red under the `strong asymmetry assumption' that all source-relays links are in increasing order of strength, while all the relays-destination links are in decreasing order; NOT CLEAR!} it was demonstrated 
and showed that a significant fraction of the capacity can be achieved by: (i) selecting a single relay, or (ii) selecting two relays and allowing them to work in a complementary fashion as in \cite{Bagheri2009}. 
Inspired by \cite{Bagheri2009}, the authors of~\cite{Fragouli2012} also showed that, for a specific HD diamond network with $N=3$ relays, at most $N+1=4$ states out of the $2^{N}=8$ possible ones are active. The authors also numerically verified that for a general HD diamond network with $N\leq 7$ relays, at most $N+1$ states are active  and conjectured that the same holds for any number of relays.

Relay networks were also studied in~\cite{OngMultiRelay}, where an iterative algorithm was proposed to determine the optimal fraction of time each HD relay transmits/receives by using DF under deterministic switching mechanism.

\subsection{Contributions}
In this work we study a general Gaussian HD relay network, whose exact capacity is unknown, by following the approach proposed in \cite{kramer-allerton}. 
Our main contributions can be summarized as follows:
\begin{enumerate}

\item We prove that NNC is optimal to within  $2.021(N+2)$ bits. %a constant gap of, uniformly over all possible channel parameters. 
This gap is smaller than the $5N$ bits gap available in the literature \cite{AvestimehrThesis,Ozgur2010} for any $N \geq 2$. 
We also show that the new gap for the HD relay network may be further decreased by considering more structured systems, such as the diamond network, for which the gap is of order $N$.

\item The bounding technique we use is tighter than the one proposed in \cite{nncLim} and, as a by-product of our approach, we reduce the gap for a general multicast complex-valued Gaussian FD network with $K$ nodes from $1.26 K$ bits to $1.021 K$ bits.

\item In order to determine the gDoF of the channel, one needs to find a tight high-SNR approximation for the different mutual information terms involved in the cut-set upper bound. As a result of independent interest,
%besides the Gaussian HD relay network studied in this paper, 
we show that such tight approximations can be found as the solution of
%We prove that the problem of finding the coefficients defining the gDoF LP is equivalent to solve  the corresponding 
Maximum Weighted Bipartite Matching (MWBM) problems, or assignment problems \cite{Burkard}. The MWBM problem is a special LP for which efficient polynomial-time algorithms, such as the Hungarian algorithm \cite{Kuhn}, exist. Although not
explored here, this technique may be useful in solving other similar problems such as that
of finding the gDoF of a general Multiple-Input-Multiple-Output (MIMO) system.
%This equivalence between the two problem gives an interesting connection between the theory of LP and graph theory.

\item We extend the results of \cite{Bagheri2009} from a 2-relay diamond network to a general 2-relay network and we show that, out of the $2^{N}=4$ possible states, at most $N+1=3$ are active. Similarly to the HD diamond network studied in \cite{Fragouli2012}, we verify through extensive numerical evaluations that, for a general relay network with $N\leq 8$ relays, at most $N+1$ states are active. Based on this evidence, we conjecture that the conjecture of \cite{Fragouli2012} holds for any HD relay network. 
%that in a general relay network with $N$ relays, there exists an optimal schedule with at most $N+1$ active states out the of the possible $2^{N}$ states. 
% This conjecture extends the one presented in \cite{Fragouli2012} valid only for the HD diamond network, and it makes use of the equivalence explained above. This conjecture is supported by the analytical proof in the case of $2$ relays, as well as by numerical simulations.

\item We finally consider a general relay network with $N=2$ relays. We highlight under which channel conditions a best-relay selection scheme is strictly suboptimal in terms of gDoF and we gain insight into the nature of the rate gain attainable in networks with multiple relays. For example, we show when the interaction between the relays, which is impossible in diamond networks, increases the gDoF.
%In other words, we show when exploiting both relays gives a higher gDoF than that attained by using only the best relay. Moreover, we compute the loss that incurs by using HD with respect to FD.

\end{enumerate}

\subsection{Paper Organization}
The rest of the paper is organized as follows. 
Section \ref{sec:sysmodl} describes the channel model and defines the gDoF and the notion of capacity to within a constant gap. 
Section \ref{sec:gapgen} shows that the cut-set upper bound and the NNC lower bound for a general Gaussian HD relay network are to within a constant gap from each another.
Section \ref{sec:optschedules} proves the equivalence between the problem of finding the coefficients of the linear inequality constraints of the LP derived from the cut-set upper bound and the MWBM problem; it also shows that, for a 2-relay network, the number of active states in the cut-set upper bound is at most 3; 
%the coefficients of the gDoF LP and the associated weighted bipartite matching problem and 
it finally presents a conjecture regarding the maximum number of active states sufficient to characterize the cut-set upper bound for a general relay network and for any number of relays.
Section \ref{sec:example} provides an example of a HD relay network with $N=2$ relays; it determines under which channel conditions the gDoF achieved with the best-relay selection strategy is strictly smaller than the gDoF attained by exploiting both relays; it provides insights into the synergies of multiple relays.
Section \ref{sec:conl} concludes the paper.

\subsection*{Notation}
We use the notation convention of \cite{book:ElGamalKim2012}: 
$[n_1:n_2]$ is the set of integers from $n_1$ to $n_2 \geq n_1$,
$[x]^+ := \max\{0,x\}$ for $x\in\mathbb{R}$;
$Y^{j}$ is a vector of length $j$ with components $(Y_1,\ldots,Y_j)$;
for an index set $\mathcal{A}$ we let $Y_{\mathcal{A}} = \{ Y_j : j\in \mathcal{A} \}$;
$\mathbf{0}_j$ is a column vector of length $j$ of all zeros;
$\mathbf{1}_j$ is a column vector of length $j$ of all ones;
$\mathbf{I}_j$ is the identity matrix of dimension $j$;
$f_1(x)\doteq f_2(x)$ means that $\lim_{x\to\infty}f_1(x)/f_2(x)=1$.
$\vert A\vert $ indicates the determinant of the matrix $A$ or the cardinality of the set $A$, which one is usually clear from the context, while $\|a\|$ is the Euclidean length of the vector $a$.
To indicate a sub matrix of the matrix $\mathbf{A}$ where only the rows indexed by the set $\mathcal{R}$
and the columns indexed by the set $\mathcal{C}$ are retained, we use the Matlab-inspired notation $\mathbf{A}_{\mathcal{R},\mathcal{C}}$. Moreover, for a square matrix $\mathbf{A}$, ${\rm diag}[\mathbf{A}]$ is a vector containing the diagonal elements of $\mathbf{A}$, while for a vector $\mathbf{a}$, ${\rm diag}[\mathbf{a}]$ is a diagonal square matrix with the elements of $\mathbf{a}$ on the main diagonal. $X \sim \mathcal{N} \left ( \mu, \sigma^2 \right )$ indicates that $X$ is a proper-complex random variable distributed normally with mean $\mu$ and variance $\sigma^2$.

\section{System model}
\label{sec:sysmodl}

%\subsection{Notation} 
\subsection{General memoryless relay network}
\label{sec:channel model:general}

A memoryless relay network has one source (node 0), one destination (node $N+1$), and $N$ relays indexed from $1$ to $N$. It consists of $ N+1$ input alphabets $\left (\mathcal{X}_1,\cdots,\mathcal{X}_{N},\mathcal{X}_{N+1} \right )$ (here $\mathcal{X}_i$ is the input alphabet of node~$i$ except for the source/node~0 where, for notation convenience, we use $\mathcal{X}_{N+1}$ rather than $\mathcal{X}_{0}$), $N+1$ output alphabets $\left (\mathcal{Y}_{1},\cdots,\mathcal{Y}_{N},\mathcal{Y}_{N+1} \right )$ (here $\mathcal{Y}_i$ is the output alphabet of node~$i$), and a transition probability $\mathbb{P}_{Y_{[1:N+1]}|X_{[1:N+1]}}$.
The source has a message $W$ uniformly distributed on $[1:2^{n R}]$ for the destination, where $n$ denotes the codeword length and $R$ the transmission rate in bits per channel use (logarithms are in base $2$). % \footnote{Logarithms are in base 2.}.
At time $i$, $i \in [1:n]$, the source maps its message $W$ into a channel input symbol $X_{N+1,i}\left ( W \right )$, and the $k$-th relay, $k\in[1: N]$, maps its past channel observations into a channel input symbol $X_{k,i}\left ( Y_{k}^{i-1} \right )$. The channel is assumed to be memoryless, that is, the following Markov chain holds for all $i\in[1:n]$ 
\begin{align*}
(W,Y_{[1: N+1]}^{i-1},X_{[1: N+1]}^{i-1}) 
\to X_{[1: N+1],i} 
\to Y_{[1: N+1],i}.
\end{align*}
%{\red SUPER WRONG!!! YOU CANNOT DROP TERMS IN THE MIDDLE OF A MAR N+2OV CHAIN!!!
%\begin{align*}
%(W,Y_{k}^{i-1},Y_ N+2^{i-1},X_1^{i-1},X_{k}^{i-1}) \to (X_{1,i},X_{k,i}) \to (Y_{k,i},Y_{ N+2,i}).
%\end{align*}}
At time $n$, the destination outputs an estimate of the message based on all its channel observations as $\widehat{W} \left (Y_{N+1}^n \right )$.
The capacity is the largest nonnegative rate such that $\mathbb{P}[\widehat{W} \!\neq\! W] \rightarrow 0$ as $n \!\rightarrow\! +\infty$. 

In this general memoryless framework, each relay can listen and transmit at the same time, i.e., it is a FD node.
HD channels are a special case of the memoryless FD framework in the following sense~\cite{kramer-allerton}.
With a slight abuse of notation compared to the previous paragraph, we let the channel input of the $k$-th relay, $k\in[1: N]$, be the pair $(X_k,S_k)$, where $X_k\in \mathcal{X}_k$ as before and $S_k\in\{0,1\}$ is the {\em state} random variable that indicates whether the $k$-th relay is in receive-mode ($S_k=0$) or in transmit-mode ($S_k=1$).
In the HD case the transition probability is specified as $\mathbb{P}_{Y_{[1:N+1]}|X_{[1:N+1]},S_{[1:N]}}$.

% By adopting this convention, there is no need to develop a separate theory for HD channels, but the HD constraints can be incorporated into the FD framework.

\subsection{The Gaussian HD relay network}
\label{sec:channel model:G-HD-RC model}

The single-antenna complex-valued power-constrained Gaussian HD relay network is described by the input/output relationship
\begin{subequations}
\begin{align}
&\mathbf{Y} =\mathbf{H}_{\rm eq} \mathbf{X} +\mathbf{Z},\\
&\mathbf{H}_{\rm eq}
:=
\begin{bmatrix}
 \mathbf{I}_N-\mathrm{diag}[\mathbf{S}] & \mathbf{0}_N \\
 \mathbf{0}_{N}^T  & 1 \\
 \end{bmatrix}
\ \mathbf{H} \
\begin{bmatrix}
 \mathrm{diag}[\mathbf{S}] &   \mathbf{0}_N \\
 \mathbf{0}_{N}^T & 1 \\
 \end{bmatrix}
\end{align} 
\label{eq:chann}
\end{subequations}
where
\begin{itemize}

\item 
$\mathbf{Y}:=[Y_{1},\ldots,Y_{N},Y_{N+1}]^T \in \mathbb{C}^{N+1}$ is the vector of the received signals.

\item 
$\mathbf{X}:=[X_{1},\ldots,X_{N},X_{N+1}]^T \in \mathbb{C}^{N+1}$ is the vector of the transmitted signals (recall that, although the source is referred to as node~0, its input is indicated as $X_{N+1}$ rather than $X_0$). Without loss of generality, we assume that the channel inputs are subject to the average power constraint $\mathbb{E} \left [ |X_{k}|^2   \right ] \leq 1$, $k \in [1: N+1]$.

\item 
$\mathbf{S}:= [S_1,\ldots,S_{N}]\in\{0,1\}^N$ is the vector of the binary relay states, which takes into account if the $k$-th relay is receiving ($S_k=0$) or transmitting ($S_k=1$) for $k\in[1: N]$.

\item 
$\mathbf{H} \in \mathbb{C}^{(N+1)\times (N+1)}$ is the constant channel matrix known by all terminals defined as
\begin{align}
\mathbf{H} :=
\begin{bmatrix}
\mathbf{H}_{\rm r \to r} & \mathbf{H}_{\rm s \to r} \\
\mathbf{H}_{\rm r \to d} & \mathbf{H}_{\rm s \to d} \\
\end{bmatrix}.
\label{eq:channel2}
\end{align}
The entry in position $(i,j)$ of the channel matrix in~\eqref{eq:channel2} represents the channel from node $j$ to node $i$, $(i,j) \in [1: N+1]^2$, in particular:
\begin{itemize}

\item
$\mathbf{H}_{\rm r \to r} \in \mathbb{C}^{N\times N}$ defines the network connections among relays, i.e., 
%is the matrix with the links between the relays, i.e., the entry $h_{ij}$ in 
$[\mathbf{H}_{\rm r \to r}]_{ij}$, $(i,j) \in [1: N]^2$, is the channel gain from the $j$-th relay to the $i$-th relay. Notice that the entries on the main diagonal of $\mathbf{H}_{\rm r \to r}$ do not matter for channel capacity, since the $k$-th relay, $k \in [1: N]$ can remove the `self-interference' $X_k$ from $Y_k$.

\item
$\mathbf{H}_{\rm s \to r} \in \mathbb{C}^{N\times 1}$ is the column vector which contains the channel gains from the source to the relays, i.e., $[\mathbf{H}_{\rm s \to r}]_{i,1}$, $i\in [1: N]$, is the channel gain from the source to the $i$-th relay;

\item 
$\mathbf{H}_{\rm r \to d} \in \mathbb{C}^{1\times N}$ is the row vector which contains the channel gains from the relays to the destination, i.e., $[\mathbf{H}_{\rm r \to d}]_{1,i}$, $i\in [1: N]$, is the channel gain from the $i$-th relay to the destination;

\item
$\mathbf{H}_{\rm s \to d} \in \mathbb{C}^{1\times 1}$is the channel gain between the source and the destination (recall that by our notation the source input is indicated as $X_{N+1}$ rather than $X_0$).
\end{itemize}

\item 
$\mathbf{Z}:=[Z_{1},\ldots,Z_{N},Z_{N+1}]^T \in \mathbb{C}^{N+1}$ is the jointly Gaussian noise vector. Without loss of generality, the noises are assumed to have zero mean and unit variance.
Furthermore we assume, not without loss of generality \cite{ZhangCorrNoise}, that the noises are independent, i.e., the covariance of $\mathbf{Z}$ is the identity matrix.

\end{itemize}

The capacity of the Gaussian HD relay network in \eqref{eq:chann} is not known in general. 
In order to evaluate the ultimate performance of this system we make use of two metrics: the gDoF and the capacity to within a constant gap. % These quantities are defined as: 
The capacity to within a constant gap is defined as:
\begin{definition}
The capacity $\mathsf{C}$ of the Gaussian HD relay network in~\eqref{eq:chann} is said to be known to within $\mathsf{GAP}$~bits if one can show an achievable rate $R^{\rm(in)}$ and an outer bound $R^{\rm(out)}$ such that
\begin{align}
R^{\rm(in)}\leq \mathsf{C} \leq R^{\rm(out)}
\leq R^{\rm(in)} + \mathsf{GAP}, %\log(2),
\label{eq:capgap}
\end{align}
where $\mathsf{GAP}$ is a constant that does not depend on the channel gain matrix $\mathbf{H}$ in \eqref{eq:chann}.
\end{definition}
Knowing the capacity to within a constant gap implies the exact knowledge of the gDoF defined as:
\begin{definition}
The gDoF of the Gaussian HD relay network in~\eqref{eq:chann} is defined as
\begin{align}
\mathsf{d} &:= \lim_{\mathsf{SNR}\to+\infty} \frac{\mathsf{C}}{\log(1+\mathsf{SNR})},
\label{eq:gDoFdefinition}
\end{align}
where $\mathsf{C}$ is the capacity and $\mathsf{SNR}\in\mathbb{R}^+$ parameterizes the channel gains as $ |h_{ij}|^2 = \mathsf{SNR}^{\beta_{ij}}$, for some non-negative $\beta_{ij}$, $(i,j)\in[1: N+1]^2$. 
\end{definition}
%
%From the two definitions above we see that 
The gDoF in \eqref{eq:gDoFdefinition} is an exact characterization of the capacity at high-SNR, while the capacity to within a constant gap in \eqref{eq:capgap} quantifies how far inner and outer bounds are in the worst $\mathsf{SNR}$ scenario.
%is a result valid at finite-SNR. In other words, the gap is the worst case capacity guarantee in the sense that it shows that inner and outer bounds are never too far away.

\section{Capacity to within a constant gap}
\label{sec:gapgen}
This section is devoted to the capacity characterization of the Gaussian HD relay network in~\eqref{eq:chann} to within a constant gap. To accomplish this, we first adapt the cut-set upper bound \cite{ KramerGastparGuptaIT05} and the NNC lower bound \cite{nncLim} to the HD case by following the approach proposed in \cite{kramer-allerton}. Then we show that these bounds are at most a constant number of bits apart. Our result is:
\begin{theorem}
\label{thm:many relays}
The cut-set upper bound for the HD relay network with $N$ relays  is achievable to within %in~\eqref{eq:chann}
\begin{align}
\label{eq:GapMultipleRelay}
\mathsf{GAP} \leq 2.021(N+2) \ \rm{bits},
\end{align}
by using as achievable scheme NNC with deterministic switch.
\end{theorem}

\begin{IEEEproof}
%%%The proof can be found in Appendix~\ref{app:prop:gap multi relay}.
Here we prove a general gap result for multicast single-antenna complex-valued power-constrained Gaussian HD networks in the spirit of \cite[Theorem 4]{nncLim}.
The channel model is defined as in the Section \ref{sec:channel model:G-HD-RC model} except that each node $k\in[1:K]$ has an independent message of rate $R_k$ { for the nodes indexed by $\mathcal{D}\not= \emptyset$} so that the channel input/output relationship reads $\mathbf{Y} = (\mathbf{I}_K - \mathbf{S}) \ \mathbf{H} \ \mathbf{S} \ \mathbf{X}+\mathbf{Z}$.
The gap for a HD relay network with $N$ relays is a special case of this setup for $K=N+2$.

%We consider a general multicast network with $K$ nodes, where each node $k \in [1:K]$ has an independent message $W_k$ for all the other nodes. This network is described by the input/output relationship
%\begin{align*}
%\mathbf{Y} = (\mathbf{I}_K - \mathbf{S})\mathbf{H}\mathbf{S}\mathbf{X}+\mathbf{Z}.
%\end{align*}
%where $\mathbf{Y} \in \mathbb{C}^{K}$ is the column vector of the received signals, $\mathbf{S} \in \mathbb{C}^{K \times K}$ is the diagonal matrix of the binary node states, $\mathbf{H} \in \mathbb{C}^{K \times K}$ is the constant channel matrix, where the entry $h_{ij}$ with $(i,j)\in [1:K]^2$ is the channel from node $j$ to node $i$, $\mathbf{X} \in \mathbb{C}^{K}$ is the column vector of the transmitted signals with $\mathbb{E} \left [ |X_{k}|^2   \right ] \leq 1$, $k \in [1:K]$ and $\mathbf{Z} \in \mathbb{C}^{K}$ is the jointly Gaussian noise column vector (noises are assumed to have zero mean and unit variance and are independent). We remark here that the multi-relay network studied in this work is a special case of this multicast network where $K=N+2$ ($N$ relays, $1$ source and $1$ destination).

The capacity of a HD Gaussian multicast network can be lower bounded by adapting the NNC scheme for the general memoryless network~\cite{nncLim} to the HD case by following \cite{kramer-allerton}. For each $\mathcal{A}\subseteq[1:K]\backslash\emptyset$ { and such that $\mathcal{A}^c \cap \mathcal{D} \not=\emptyset$}, similarly to~\cite[eq.(20)]{nncLim}, the NNC lower bound  gives
%\begin{subequations}
\begin{align}
\sum_{i\in\mathcal{A}}R_i
&\geq 
 I(X_{\mathcal{A}}; \widehat{Y}_{\mathcal{A}^c}| X_{\mathcal{A}^c},S_{[1: K]})  %\nonumber
-I(Y_{\mathcal{A}};\widehat{Y}_{\mathcal{A}}| \widehat{Y}_{\mathcal{A}^c},X_{[1: K]},S_{[1: K]})  \nonumber
\\&\geq
\sum_{s=1}^{2^{K}}\lambda_{s} \ \log\left|\mathbf{I}_{|\mathcal{A}^c|} + \frac{1}{1+\sigma^2}\ \mathbf{H}_{\mathcal{A},s}  \mathbf{H}_{\mathcal{A},s}^H\right|
-|\mathcal{A}| \log\left(1+\frac{1}{\sigma^2}\right), %\nonumber
\label{eq:eqnewmultilow}
\end{align}
%\end{subequations}
where $\lambda_{s}:= \mathbb{P}[S_{[1: K]}=s] \in[0,1], \forall s  \in  [1:2^{K}] : \ \sum_{s=1}^{2^{K}}\lambda_{s}   =   1$, where ``$S_{[1: K]}=s$'' is a shorthand notation for $S_{[1: K]} = [S_1, \ldots, S_K]$
%that the $j$-th entry of $S_{[1: K]}$ is equal to the $j$-th digit in the binary expansion of the number $s-1$, $j \in [1: K]$; for example, 
where $S_k\in[0:1], \ \forall k\in[1:K],$ are such that
$s = 1+\sum_{k=1}^{K} S_k 2^{K-k}$
(for example ``$S_{[1: 5]}= 8$'' means $S_{[1:5]}=[S_1,\ldots,S_5]=[0,0,1,1,1]$ since $8-1=0 \cdot 2^4 + 0 \cdot 2^3+ 1\cdot 2^2+ 1 \cdot 2^1 + 1 \cdot 2^0$; similarly ${\rm diag}[s]|_{s=8} = {\rm diag}\big[[0,0,1,1,1]\big]$).
The matrix $\mathbf{H}_{\mathcal{A},s}$ is defined as
$\mathbf{H}_{\mathcal{A},s}:=\big[ (\mathbf{I}_K-{\rm diag}[s])\ \mathbf{H} \ {\rm diag}[s] \big]_{\mathcal{A}^c,\mathcal{A}}$.
In all states $s\in  [1:2^{K}]$, we consider i.i.d. $\mathcal{N}   \left( 0,1 \right)$ inputs, time sharing random variable $Q$ set to $Q  =  S_{[1: K]}$ (with this choice the nodes can coordinate), and compressed output $\widehat{Y}_k   :=   Y_k   +   \widehat{Z}_k$, $k \in [1:K]$, for $\widehat{Z}_k  \sim  \mathcal{N}(0,\sigma^2)$ independent of all other random variables (note that the variance of $\widehat{Z}_k$ does not depend on the user index $k$).

The cut-set upper bound in \cite{ KramerGastparGuptaIT05} adapted to the HD case \cite{kramer-allerton} gives, similarly to~\cite[eq.(19)]{nncLim}, 
%\begin{subequations}
\begin{align}
\sum_{i\in\mathcal{A}}R_i
&\leq 
 I(X_{\mathcal{A}},S_{\mathcal{A}}; Y_{\mathcal{A}^c}|X_{\mathcal{A}^c},S_{\mathcal{A}^c}) \nonumber
\\& \leq I(S_{\mathcal{A}}; Y_{\mathcal{A}^c}) + I(X_{\mathcal{A}}; Y_{\mathcal{A}^c}|X_{\mathcal{A}^c},S_{[1:K]}) \nonumber
\\& \leq
   H(S_{\mathcal{A}})
+ \sum_{s=1}^{2^{K}}\lambda_{s} \ \log\left|\mathbf{I}_{|\mathcal{A}^c|} + \mathbf{H}_{\mathcal{A},s} \mathbf{K}_{\mathcal{A},s} \mathbf{H}_{\mathcal{A},s}^H\right| \nonumber
\\&\stackrel{\rm(a)}{\leq}
|\mathcal{A}|\log(2)
+\sum_{s=1}^{2^{K}}\lambda_{s} \ \log\left|\mathbf{I}_{|\mathcal{A}^c|} + \frac{1}{1+\sigma^2}\mathbf{H}_{\mathcal{A},s}\mathbf{H}_{\mathcal{A},s}^H\right| \nonumber
 \\& \quad+\sum_{s=1}^{2^{K}}\lambda_{s} {\rm Rank}[\mathbf{H}_{\mathcal{A},s}]
 \log\left( \eu \max\left \{1,\frac{1+\sigma^2}{\eu} \ \frac{|\mathcal{A}|}{{\rm Rank}[\mathbf{H}_{\mathcal{A},s}]}\right \}\right)^{\min\left \{\frac{1+\sigma^2}{\eu},\frac{{\rm Rank}[\mathbf{H}_{\mathcal{A},s}]}{|\mathcal{A}|}\right\}} \nonumber
\\& \stackrel{\rm(b)}{\leq} |\mathcal{A}|\log(2)
+\sum_{s=1}^{2^{K}}\lambda_{s} \ \log\left|\mathbf{I}_{|\mathcal{A}^c|} + \frac{1}{1+\sigma^2}\mathbf{H}_{\mathcal{A},s}\mathbf{H}_{\mathcal{A},s}^H\right| \nonumber
\\&  \quad+\!\min \{|\mathcal{A}|,|\mathcal{A}^c|\}
\log\left(\eu \max\left \{1,\frac{1\!+\!\sigma^2}{\eu} \ \frac{|\mathcal{A}|}{\min \{|\mathcal{A}|,|\mathcal{A}^c|\}}\right \}\right)^{\min\left \{\frac{1\!+\!\sigma^2}{\eu},\frac{\min \{|\mathcal{A}|,|\mathcal{A}^c|\}}{|\mathcal{A}|}\right\}},
\label{eq:eqnewmulti}
\end{align}
%\end{subequations}
for all $\mathcal{A}\subseteq[1:K]\backslash\emptyset$ { and such that $\mathcal{A}^c \cap \mathcal{D} \not=\emptyset$ (see ~\cite[eq.(4)]{nncLim})}, where $\mathbf{K}_{\mathcal{A},s}$ represents the covariance matrix of $X_{\mathcal{A}}$ conditioned on $S_{[1: K]}=s$. The inequality in (a) follows since the entropy of a random variable can be upper bounded with the support of the variable and by using \cite[Lemma 1]{nncLim} for some $\sigma^2 \geq \eu -2$. The inequality in (b) is due to the fact that the function is increasing with respect to the rank of the channel matrix and the rank is upper bounded by the minimum between the number of rows and of columns.

By letting $\gamma=1+\sigma^2 \geq \eu -1$ and $\mu=\frac{|\mathcal{A}|}{K}$, the gap between the cut-set upper bound in~\eqref{eq:eqnewmulti} and the NNC lower bound in~\eqref{eq:eqnewmultilow} becomes
\begin{align*}
\mathsf{GAP}
& \leq K \left ( \min_{\gamma \geq \eu-1} \max_{\mu\in[0:1]} \left \{ \mu\log\left(\frac{2\gamma}{\gamma-1}\right) \right. \right. +
\\& \left. \left. + \min \left \{\mu,1-\mu \right \}\min\left \{\frac{\gamma}{\eu},\frac{\min \{\mu,1-\mu\}}{\mu}\right\}  
\log\left(\max\left \{\eu,\frac{\gamma \mu}{\min \{\mu,1-\mu\}}\right \}\right)\right\}
\right)
\\& \stackrel{\rm(c)}{=}\frac{K}{2} \left ( \min_{\gamma \geq \eu-1} \left \{\log\left(\frac{2\gamma}{\gamma-1}\right) 
+ \min\left \{\frac{\gamma}{\eu},1\right\}  
\log\left(\eu \max\left \{1,\frac{\gamma}{\eu} \right \} \right)%\left. \right |_{x:=\frac{\gamma}{\eu}}
\right \}
\right )
\\& \stackrel{\rm(d)}{=} 2.021 K \ \text{bits},
\end{align*}
where the maximum over $\mu \in [0,1]$ is attained at $\mu^{\rm{opt}}=\frac{1}{2}$ and the minimum over $\gamma$ is attained at $\gamma^{\rm{opt}}=\frac{\sqrt{4\eu+1}+1}{2}$ and these results lead to the equalities in (c) and (d),  respectively. 
%in order to obtain the special case of the multi-relay network we get \eqref{eq:GapMultipleRelay}.
%{\red TO BE CONTINUED}
\end{IEEEproof}

%%% NO LONGER NEEDED
%%%For $N\gg1$, the optimal value of $\ell$ in \eqref{eq:GapMultipleRelay} is $\ell \cong \frac{N}{2}$ and the corresponding gap becomes 
%%%\begin{align}
%%%\mathsf{GAP} \cong \frac{N+2}{2} \log\Big(4(N+2)\Big).
%%%\label{eq:multirelaylimit}
%%%\end{align}
%%%Fig.~\ref{fig:GapVaryingUsers} shows the gap in \eqref{eq:GapMultipleRelay} (blue curve) as a function of the number of relays $N$, as well as its the limit behavior in \eqref{eq:multirelaylimit} (red curve) and the previously known gap (green curve) from \cite{Ozgur2010}. 
%%%%We observe that the gap in \cite{Ozgur2010}, obtained by assuming a random switch at the relays and by using a scheme based on nested lattice codes, is bigger than the one computed here for any value of $ N+2$.

We observe that the gap in \eqref{eq:GapMultipleRelay} improves, for any number of relays greater than one, on the previously known gap result of $5N \ \rm{bits}$ \cite{Ozgur2010}. Moreover:

\begin{remark}
The gap result in \eqref{eq:GapMultipleRelay} for $N\!=\!1$ gives $\mathsf{GAP}\!\leq \!6.0630$~bits, which is greater than the 1.61~bits gap we found for the single relay case in \cite{cardoneISIT2013}. This is due to the fact that the bounding steps in the special case of $N\!=\!1$ are tighter than those we used here for a general $N$.

Note also that for a single relay, PDF is optimal to within 1 bit \cite{cardoneISIT2013}. However, PDF does not seem to be easily extendable to networks with an arbitrary number of relays \cite{KramerGastparGuptaIT05}, which is the main motivation for considering NNC here. 
\end{remark}

\begin{figure*}
\centering
\includegraphics[width=0.7\textwidth]{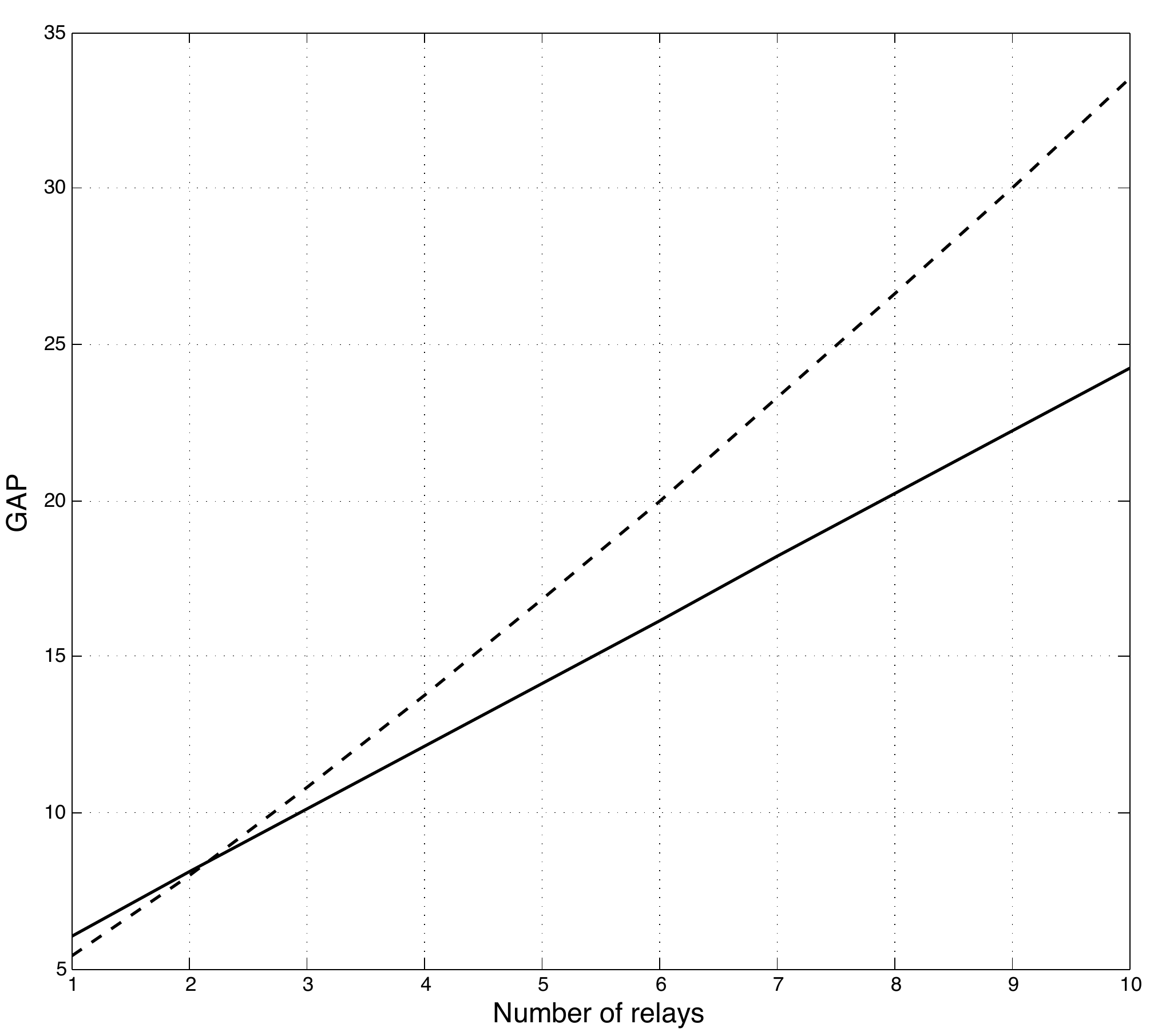}
\caption{Gap in \eqref{eq:GapMultipleRelay} (solid curve) and gap in \eqref{eq:oldgap} (dashed curve) for the HD Gaussian relay network. The gap in \eqref{eq:GapMultipleRelay} is smaller than that in \eqref{eq:oldgap} for any number of relays greater than or equal to $2$.  
%{\red remove legend; use black color with 2 different line styles; add description here.}
}
\label{fig:gapNewOld}
\end{figure*}

\begin{remark}
 In a preliminary version of this work \cite{USITW2013multirelay}, by using a bounding technique as in \cite[pages 20-5, 20-7]{ElGamalLecture} we obtained
\begin{align}
\mathsf{GAP}\cong \frac{N+2}{2} \log \left( 4(N+2) \right) \ \rm{bits}.
\label{eq:oldgap}
\end{align}
As shown in Fig. \ref{fig:gapNewOld} the gap in \eqref{eq:GapMultipleRelay} is smaller that the one in \eqref{eq:oldgap} for $N\geq 2$. This is accomplished thanks to the tighter bound from \cite[Lemma 1]{nncLim}.
\end{remark}

\begin{remark}
From \cite{nncLim}, the gap of a general $K$-user multicast complex-valued FD Gaussian network is $(2 \times 0.63)K$. However, by using the tighter bound $ {\rm Rank}[\mathbf{H}_{\mathcal{A},s}] \leq \min \{|\mathcal{A}|,|\mathcal{A}^c|\}$, instead of $ {\rm Rank}[\mathbf{H}_{\mathcal{A},s}] \leq  |\mathcal{A}|$, the gap can be reduced to $(2 \times 0.5105)K$.
\label{rem:remark3}
\end{remark}

A smaller gap than the one in \eqref{eq:GapMultipleRelay} may be obtained by deriving tighter bounds on specific network topologies.
%with more sophisticated bounding techniques or for more structured networks.
%This can be accomplished by several means.  
For example,
%%%\begin{itemize}
%%%
%%% NO LONGER NEEDED
%%%\item 
%%%By using more complex and sophisticated bounding techniques: in~\cite{nncLim} an upper bound on the water-filling power allocation for a general Multiple-Input-Multiple-Output (MIMO) channel was derived. This more involved upper bound could be used here to obtain a smaller gap. We note that our bounding technique applied to the Gaussian FD relay network gives a gap of $ \frac{N+2}{2}\log\Big(2(N+2)\Big)$, which is larger than $0.63(N+2)$ found in~\cite{nncLim}. Our bound uses the well known fact that for a positive semidefinite matrix $\mathbf{K}$ we have $\mathbf{K} \preceq {\rm Trace}[\mathbf{K}] \ \mathbf{I}$. 
%%%
%%%\item
in~\cite{arXiv:1207.5660} it was found that for a Gaussian FD diamond network with $N$ relays the gap is of the order $\log(N)$, rather than linear in $N$~\cite{nncLim}. Moreover, for a symmetric FD diamond network with $N$ relays the gap does not depend on the number of relays and it is upper bounded by 2~bits \cite{NiesenDiggavi}. The key difference between a general relay network and a diamond network is that for each subset $\mathcal{A}$ we have ${\rm Rank}[\mathbf{H}_{\mathcal{A}}] \leq 2$, i.e., the rank of any channel sub-matrix does no longer depend on the cardinality of the index set $\mathcal{A}$.  Based on the simpler topology of a diamond network we have:
\begin{proposition}\label{prop:hd diamond}
The cut-set upper bound for the Gaussian HD diamond network with $N$ relays is achievable to within 
\begin{align}
\mathsf{GAP} \leq
\frac{4}{N+2} \log \left( \frac{N+2}{2}+\frac{(N+2)^3}{8}\right)+(N+2)\log \left( 2+\frac{8}{(N+2)^2} \right)
\label{eq:gap diamond}
\end{align}
bits.
\end{proposition}
\begin{IEEEproof}
The proof can be found in Appendix~\ref{app:hd diamond}.
\end{IEEEproof}
When $N\gg1$, the gap in \eqref{eq:gap diamond} can be approximated as
$\mathsf{GAP} \cong N \ \rm{bits}$.
As expected, the gap in \eqref{eq:gap diamond} for the HD diamond network is in general (for $N \geq 3$) 
%{\blue does it make sense? the fact that for $N=2$ it is not smaller could be due to the fact that we used the bound rank less than 2?}
smaller than that in \eqref{eq:GapMultipleRelay} computed for the general HD relay network; this is in line with what happens in the FD case. However, in FD for the diamond network the gap is logarithmic in $N$ \cite{arXiv:1207.5660}, while the gap in \eqref{eq:gap diamond} still grows linearly with $N$. This is in part %{\blue only in part? since as $N$ grows the gap is of $N$ should not be this is entirely due?} 
due to the fact that, in the HD outer bound, there is an entropy term due to the random switch that is maximized by considering a uniform probability over the all possible listen / transmit states that, for the multicast network, are $2^{N+2}$ (see Appendix~\ref{app:hd diamond}). 

As we shall see in the next section, for a general HD network with $N$ relays, only $N+1$ states, out of the possible $2^N$ states, appear to be needed to characterize the cut-set upper bound. It is subject of current investigation on how to use this observation to develop bounds leading to a smaller gap.
%A possible way to further decrease this gap would be to develop an achievable scheme that exploits the random switch at the relays to convey information.
%%%\end{itemize}

\section{Analysis of the optimal schedule}
\label{sec:optschedules}
In general, for a $N$-relay network, $2^{N}$ states are possible. A capacity achieving scheme must optimize the fraction of time each of these states occurs. In \cite{Bagheri2009}, it was proved that for a diamond network with $N=2$ relays, out of the $2^{N}=4$ possible states, at most $N+1=3$ have a non-zero probability and are sufficient to characterize the cut-set upper bound, i.e., we say that there are $N+1=3$ {\em active states}. In~\cite{Fragouli2012}, the authors extended the result of \cite{Bagheri2009} to a special case of diamond network with $N=3$ relays; based on numerical evidences, \cite{Fragouli2012} conjectured that for a $N$-relay diamond network out of the $2^{N}$ possible states at most $N+1$ states are active.
% / have a non-zero probability.
%, by numerically showing that this conjecture holds for $ N+2-2 \leq 7$. 
Here we extend these results to a general Gaussian HD relay network as follows. The claim ``out of $2^{N}$ possible states only $N+1$ states are active as far as gDoF is concerned'' is proved analytically for $N=2$, shown to hold by numerical evaluations for $N\leq 8$ and conjectured to hold for any $N$. If the conjecture were true, it would show that HD relay networks have intrinsic properties regardless of their topology, i.e., known results for diamond networks are not a consequence of the simplified network topology.

In order to determine the gDoF we must find a tight high-SNR approximation for the different MIMO-type mutual information terms involved in the cut-set upper bound (see Section~\ref{sec:gapgen} eq.\eqref{eq:eqnewmulti}). 
As a result of independent interest, besides the application to the Gaussian HD relay network studied in this paper, we first show that such an approximation can be found as the solution of a Maximum Weighted Bipartite Matching (MWBM) problem.

\subsection{The maximum weighted bipartite matching (MWBM) problem}
\label{sec:wbmp}
%Here we formally prove that, for a general relay network (with any number of relays), the problem of finding the coefficients characterizing the gDoF expression is equivalent to solve the associated weighted bipartite matching problem. We start by defining some concepts from graph theory that will be useful next.
%\begin{definition}
In graph theory, a weighted bipartite graph, or bigraph, is a graph whose vertices can be separated into two sets such that each edge in the graph has exactly one endpoint in each set. Moreover, a non-negative weight is associated with each edge in the bigraph \cite{Lovasz}. A matching, or independent edge set, is a set of edges without common vertices \cite{Lovasz}. 
%\end{definition}
%
%\begin{definition}
%In graph theory, a
The MWBM problem, or assignment problem, is defined as a matching  where the sum of the edge weights in the matching has the maximal value \cite{Burkard}.
%\end{definition}
%
%\begin{definition}
The Hungarian algorithm  is a polynomial time algorithm that efficiently solves the assignment problem \cite{Kuhn}.
%\end{definition}
%
Equipped with these definitions, we now show the following high-SNR approximation of the MIMO capacity:
%For completeness we report the following definition.
%A combination is a way of selecting elements out of a set, where (unlike permutations) order does not matter.
%A $k$-combination of a set is a subset of $k$ distinct elements of the set.
%If the set has $n$ elements the number of $k$-combinations is equal to the binomial coefficient.
%, i.e., sequences of a fixed length $k$ of elements taken from a given set of size $n$, also known as partial permutations or as sequences without repetition.
\begin{theorem}
\label{th:MWBM}
Let $\mathbf{H} \in \mathbb{R}^{ k \times n}$ be a full-rank matrix, where without loss of generality $k\leq n$.  
Let $\mathcal{S}_{n,k}$ be the set of all $k$-combinations of the integers in $[1:n]$ and $\mathcal{P}_{n,k}$ be the set of all $k$-permutations of the integers in $[1:n]$.
Then,
\begin{align}
|\mathbf{I}_k+\mathbf{H}\mathbf{H}^H|
& = 
 \sum_{\varsigma \in \mathcal{S}_{n,k}} 
 \sum_{\pi \in \mathcal{P}_{n,k}} 
\underbrace{\prod_{i=1}^{k}|h_{i,\pi(i)}|^2}_{ = \mathsf{SNR}^{\sum_{i=1}^{k}[\mathbf{B}_\varsigma]_{i,\pi(i)}} }  + \quad T \doteq \mathsf{SNR}^{\text{MWBM}(\mathbf{B})},
% 
% }_{ \doteq \mathsf{SNR}^{\max_{\pi \in \mathcal{P}_{n,k}} \sum_{i=1}^{k}\left[\beta_\varsigma\right ]_{i,\pi(i)}}}
\\&
\text{MWBM}(\mathbf{B}):={\max_{\varsigma \in \mathcal{S}_{n,k}}\max_{\pi \in \mathcal{P}_{n,k}} \sum_{i=1}^{k}\left[\mathbf{B}_\varsigma\right ]_{i,\pi(i)}},
 \label{eq:daniela}
\end{align}
where $\mathbf{B}$ is the SNR-exponent matrix defined as $[\mathbf{B}]_{ij} = \beta_{ij}\geq 0 : |h_{ij}|^2 = \mathsf{SNR}^{\beta_{ij}}$,  
$\mathbf{B}_\varsigma$ is the square matrix obtained from $\mathbf{B}$ by retaining all rows and those columns indexed by $\varsigma$, and 
$T$ is the sum of terms that overall have an exponential behavior that is less than $\text{MWBM}(\mathbf{B})$. \end{theorem}
\begin{IEEEproof}
The proof can be found in Appendix~\ref{app:MWBM}.
The expression in~\eqref{eq:daniela} is a possible way of writing the MWBM problem.
\end{IEEEproof}

Theorem~\ref{th:MWBM} establishes an interesting connection between the capacity of a MIMO channel (with independent inputs) and graph theory. Note that the high-SNR expression found in Theorem~\ref{th:MWBM} holds for correlated inputs as well, as long as the average power constraint is a finite constant.
More importantly, Theorem~\ref{th:MWBM} allows to move from DoF, where all exponents $\beta_{ij}$ have the same value, to gDoF, where different channel gains have different exponential behavior. 
DoF is essentially a characterization of the rank of the channel matrix; gDoF captures the potential advantage due to `asymmetric' channel gains.
gDoF, to the best of our knowledge, has been investigated so far only for Single-Input-Single-Output (SISO) networks with very few number of nodes; we believe that the reason is that in these cases one has only to consider equivalent Multiple-Input-Single-Output (MISO) and Single-Input-Multiple-Output (SIMO) channels, or to explicitly deal with determinants of matrices with small dimensions. Our result extends the gDoF analysis to any MIMO channel as we explain through some examples:
\begin{enumerate}

\item
Case $k=1 \leq n$: In a MISO or SIMO channel, with channel vector $\mathbf{h} := [h_1,\ldots,h_n]$ such that $|h_k|^2=\mathsf{SNR}^{\beta_k}, k\in[1:n]$, one trivially has
\[
\log(1+\vert| \mathbf{h} \vert|^2) 
= \log\left(1+\sum_{i=1}^{n} \mathsf{SNR}^{\beta_i}\right)
\stackrel{\mathsf{SNR} \gg 1}{\doteq} \log\left(\mathsf{SNR}^{\max_{i\in[1:n]} \{ \beta_i\} }\right).
\] 
The corresponding MWBM problem has one set of vertices $\mathcal{A}_1$ consisting of $k=|\mathcal{A}_1|=1$ node and the other set of vertices $\mathcal{A}_2$ consisting of $n=|\mathcal{A}_2| \geq 1$ nodes. The weights of the edges connecting the single vertex in $\mathcal{A}_1$ to the $n$ vertices in $\mathcal{A}_2$ can be represented as the non-negative vector $\mathbf{B}=[\beta_1,\ldots,\beta_n]$. Clearly, the optimal $\text{MWBM}(\mathbf{B})=\max_{i\in[1:n]} \{ \beta_i\}$ assigns the single  vertex in $\mathcal{A}_1$ to the vertex in $\mathcal{A}_2$ that is connected to it through the edge with the maximum weight.

\item
Case $k=n=2$: As another example from the 2-user interference channel literature, consider the cut-set sum-rate upper bound~\cite{PVIT11} 
\begin{align*}
&\log\left(
\mathbf{I}_2
+
\mathbf{H}
\mathbf{H}^H
\right)
\stackrel{\mathsf{SNR} \gg 1}{\doteq} 
\log\left(\mathsf{SNR}^{\max\{\beta_{13}+\beta_{24}, \beta_{23}+\beta_{14}\}}\right),
\\&
\quad
\mathbf{H}
: = 
\begin{bmatrix}
h_{13} & h_{23}  \\
h_{14} & h_{24}  \\ 
\end{bmatrix}
= 
\begin{bmatrix}
\mathsf{SNR}^{\beta_{13}/2}\ \eu^{\jj \theta_{13}} & \mathsf{SNR}^{\beta_{23}/2}\  \eu^{\jj \theta_{23}} \\
\mathsf{SNR}^{\beta_{14}/2}\ \eu^{\jj \theta_{14}} & \mathsf{SNR}^{\beta_{24}/2}\  \eu^{\jj \theta_{24}} \\ 
\end{bmatrix}.
\end{align*}
The corresponding MWBM problem has one set of vertices $\mathcal{A}_1$ consisting of $k=|\mathcal{A}_1|=2$ nodes (for future references let us refer to these vertices as nodes 1 and 2 -- see first subscript in the channel gains)
and the other set of vertices $\mathcal{A}_2$ consisting also of $n=|\mathcal{A}_2|=2$ nodes (for future references let us refer to these vertices as nodes 3 and 4 -- see second subscript in the channel gains). The weights of the edges connecting the vertices in $\mathcal{A}_1$ to the vertices in $\mathcal{A}_2$ can be represented as the non-negative weights $\beta_{ji}, \ i=3,4, \ j=1,2 $. In this example, one possible matching assigns node 1 to node 3 and node 2 to node 4 (giving total weight $\beta_{13}+\beta_{24}$), while the other possible matching assigns node 2 to node 3 and node 1 to node 4 (giving total weight $\beta_{23}+\beta_{14}$); the best assignment is the one that gives the largest total weight.

Notice that the MWBM is a tight approximation of the $2\times 2$ MIMO capacity only when the channel matrix is full rank, see \cite[eq.(5) 1st line]{PVIT11}, but it is loose when the channel matrix is rank deficient, see \cite[eq.(5) 2nd line, and compare with eq.(11)]{PVIT11}. The reason is that the MWBM can not capture the impact of phases in MIMO situations. To exclude the case of a rank deficient channel matrix from our general setting for any value of $k$ and $n$, we may proceed as in \cite[page 2925]{WangTse}. Namely, we pose a reasonable distribution, such as for example the i.i.d. uniform distribution, on the phases  $\theta_{ji}, \ i=3,4, \ j=1,2 $, so that almost surely the channel matrix is full rank.

\item
Case $k=2, n=3$: The MWBM allows to find the high-SNR approximation of any MIMO system capacity. As an example, which to the best of our knowledge is not known from the literature, consider a full-rank MIMO systems $n=3$ transmit antennas and $k=2$ receive antennas and with SNR-exponent matrix 
$\mathbf{B}=\begin{bmatrix}
\beta_{11} & \beta_{12} & \beta_{13}  \\
\beta_{21} & \beta_{22} & \beta_{23}  \\ 
\end{bmatrix}$.
In this case we have %the MWBM gives the following high-SNR exponent for the capacity
\[
\text{MWBM}(\mathbf{B}) = \max\Big\{
\beta_{11}+\beta_{22},
\beta_{11}+\beta_{23},
\beta_{12}+\beta_{21},
\beta_{12}+\beta_{23},
\beta_{13}+\beta_{21},
\beta_{13}+\beta_{22} \Big\},
\]
which can also be obtained by tedious direct computation of the limiting value of the corresponding log-det formula.

%and the corresponding associated weighted bipartite matching problem, that can be easily solved  by applying the Hungarian algorithm \cite{ N+2uhn}. This approach provides an efficient automatic way to find the constant coefficients characterizing the gDoF expression. Once these coefficients are determined the gDoF LP can be efficiently solved numerically.
\end{enumerate}

\subsection{The gDoF for a general $N$-relay network}
With Theorem~\ref{th:MWBM} we can now express the gDoF $\mathsf{d}$ in \eqref{eq:gDoFdefinition} of the Gaussian HD relay network in \eqref{eq:chann} 
%(see Appendix~\ref{app:prop:gap multi relay} eq.\eqref{eq:eqnewmulti mimo} and eq.\eqref{eq:eqnewmultilow mimo}),  
as a LP. In particular, let
\begin{subequations}
\begin{align}
   \mathbf{f}^T &:= [\mathbf{0}^T_{2^{N}}, 1]
\\ \mathbf{x}^T &:= [\lambda_{1},\ldots, \lambda_{2^{N}},\mathsf{d}],
\end{align}
then
\begin{align}
\mathsf{d}
  &= \max \{ \mathbf{f}^T \ \mathbf{x} \} \label{eq:LP obj}
\\& \st \begin{bmatrix}
- \mathbf{A} & \mathbf{1}_{2^{N}} \\
\mathbf{1}_{2^{N}}^T & 0 \\
\end{bmatrix} \mathbf{x} \leq \mathbf{f}, \quad \mathbf{x}\geq 0, \label{eq:LP constaints}
\end{align}
where the non-negative matrix $\mathbf{A}\in\mathbb{R}^{2^N \times 2^N}$ has entries (recall that, although the source is referred to as node~0, its input is indicated as $X_{N+1}$ rather than $X_0$)
{%\red
\begin{align}
[\mathbf{A}]_{ij}
  &:= \lim_{\mathsf{SNR}\to+\infty} 
  \frac{I(X_{\mathcal{A}_i\cup\{N+1\}}; Y_{\mathcal{A}_i^c\cup\{N+1\}}|X_{\mathcal{A}_i^c},S_{[1:N]} = s_j)}{\log(1+\mathsf{SNR})}.
  \label{eq:LP coeffs as lim}
%\nonumber\\&= \text{MWBM}(\mathbf{B}_{\mathcal{A}_i,s_j}), 
% = [\mathbf{A}]_{ji}
\end{align}
}
\label{eq:LP}
\end{subequations}
By a simple application of Theorem~\ref{th:MWBM} we have
\begin{theorem}
\label{th:coeff const of LP}
%\begin{align}
$[\mathbf{A}]_{ij} = \text{MWBM}(\mathbf{B}_{\mathcal{A}_i,s_j})$.
%  \label{eq:LP coeffs with lim evaluated}
%\end{align}
\end{theorem}
The notation in eq.\eqref{eq:LP coeffs as lim} and in Theorem~\ref{th:coeff const of LP} is as follows.
$\mathbf{B}$ indicates the SNR-exponent matrix defined as $[\mathbf{B}]_{ij} = \beta_{ij}\geq 0 : |h_{ij}|^2 = \mathsf{SNR}^{\beta_{ij}}$ (defined in \eqref{eq:channel2}), and the indices $(i,j)$ have the following meaning. Index $i$ refers to a ``cut'' in the network and index $j$ to a ``state of the relays''. Both indices range in $[1:2^{N}]$ and must be seen as the decimal representation of a binary number with $N$ bits. $\mathcal{A}_i$, $i\in[1:2^{N}]$, is the set of those relays who have a one in the corresponding binary representation of $i-1$ (example for $N=3$: for $i-1 = 6 = 1 \cdot 2^2 + 1 \cdot 2^1 + 0 \cdot 2^0$ we have $\mathcal{A}_{7} = \{1,2\}$ and therefore $\mathcal{A}_{7}^c = \{3\}$).  $s_j$, $j\in[1:2^{N}]$, sets the state of a relay to the corresponding bit in the binary representation of $j-1$ (example for $N=3$: for $j-1=6$ we have $s_{7}=[1,1,0]$, which means that relays 1 and 2 are transmitting and relay 3 is receiving). Finally,
\begin{align*}
\mathbf{B}_{\mathcal{A}_i,s_j}:= \left [
\begin{bmatrix}
 \mathbf{I}_N-\mathrm{diag}[s_j] & \mathbf{0}_{N} \\
  \mathbf{0}_{N}^T & 1 \\
 \end{bmatrix}
\ \mathbf{H} \
\begin{bmatrix}
 \mathrm{diag}[s_j] & \mathbf{0}_{N}\\
 \mathbf{0}_{N}^T & 1 \\
 \end{bmatrix}
\right ]_{\{ N+1\}\cup\mathcal{A}_i^c,\{N+1\}\cup\mathcal{A}_i}.
\end{align*}
% $(\mathbf{I}-\mathbf{S}) \mathbf{B} \mathbf{S}$ by retaining the rows indexed by $\{ N+1\}\cup\mathcal{A}_i^c$ and the columns indexed by $\{N+1\}\cup\mathcal{A}_i$ for ${\rm diag}[\mathbf{S}]=S_{[1: N+1]}$ such that $[S_{[1: N]}=s_j, \ S_{N+1}=1]$.

One interesting question is how many $\lambda_{j}$, i.e., $\lambda_{j}$ is the fraction of time the network is in state $j\in[1:2^{N}]$, are strictly positive~\cite{Bagheri2009,Fragouli2012}. 
In \cite{Bagheri2009}, the authors analyzed the diamond network with $2$ relays and showed that out of the $4$ possible states only $3$ states are active. The proof considers the dual of the LP in~\eqref{eq:LP}. Here we extend the result of \cite{Bagheri2009} to the fully-connected HD relay network with $2$ relays; our proof identifies the channel conditions under which setting the probability of one of the states to zero is without loss of optimality. We have:
\begin{theorem}
\label{thm:active2relays}
For a general HD relay network with 2 relays, there exists an optimal schedule that optimizes $\mathsf{d}$ in \eqref{eq:LP obj} with at most $3$ active states.
\end{theorem}
\begin{IEEEproof}
The proof can be found in Appendix~\ref{app:hdactive2}, which uses the notation in \eqref{eq:par2rel} where:
$\alpha_{si}$ is the SNR-exponent on the link from the source to relay~$i$, $i\in[1:2]$,
$\alpha_{id}$ is the SNR-exponent on the link from relay~$i$, $i\in[1:2]$, to the destination,
$\beta_i$ is the SNR-exponent on the link from relay~$j$ to relay~$i$, $(i,j)\in[1:2]^2$ with $j\not= i$, and
the direct link from the source to the destination has SNR-exponent normalized to $1$.% without loss of generality.
\end{IEEEproof}

%The above LP, similarly to the HD diamond network case in~\cite{Bagheri2009,Fragouli2012}, where the coefficients of the linear inequality constraints can be found as the solution of certain MWBM problems.
%the coefficients of the asymptotic ($\mathsf{SNR}\rightarrow \infty$) expression of the capacity of a general relay network is equivalent to solve the assignment problem for a weighted bipartite graph where in one set we have the signals transmitted ($X$) and in the other set we have the signals received ($Y$) and where the edges have a weight given by the SNR exponent of the corresponding link. Our main result is 

%The theorem above establishes an interesting connection between the gDoF LP and the corresponding associated weighted bipartite matching problem, that can be easily solved  by applying the Hungarian algorithm \cite{Kuhn}. This approach provides an efficient automatic way to find the constant coefficients characterizing the gDoF expression. Once these coefficients are determined the gDoF LP can be efficiently solved numerically.

We conjecture that for a general HD relay network with any number of relays Theorem~\ref{thm:active2relays} continues to hold,
% there always exists an optimal schedule with a linear, instead of exponential, number of states. This
similarly to the conjecture presented in~\cite{Fragouli2012} for the diamond network. Namely:

{\bf Conjecture.} 
{\em 
For a general HD relay network with $N$ relays, there always exists an optimal schedule that maximizes the gDoF with at most $N+1$ active states. 
}

\begin{figure*}
\centering
\includegraphics[width=0.8\textwidth]{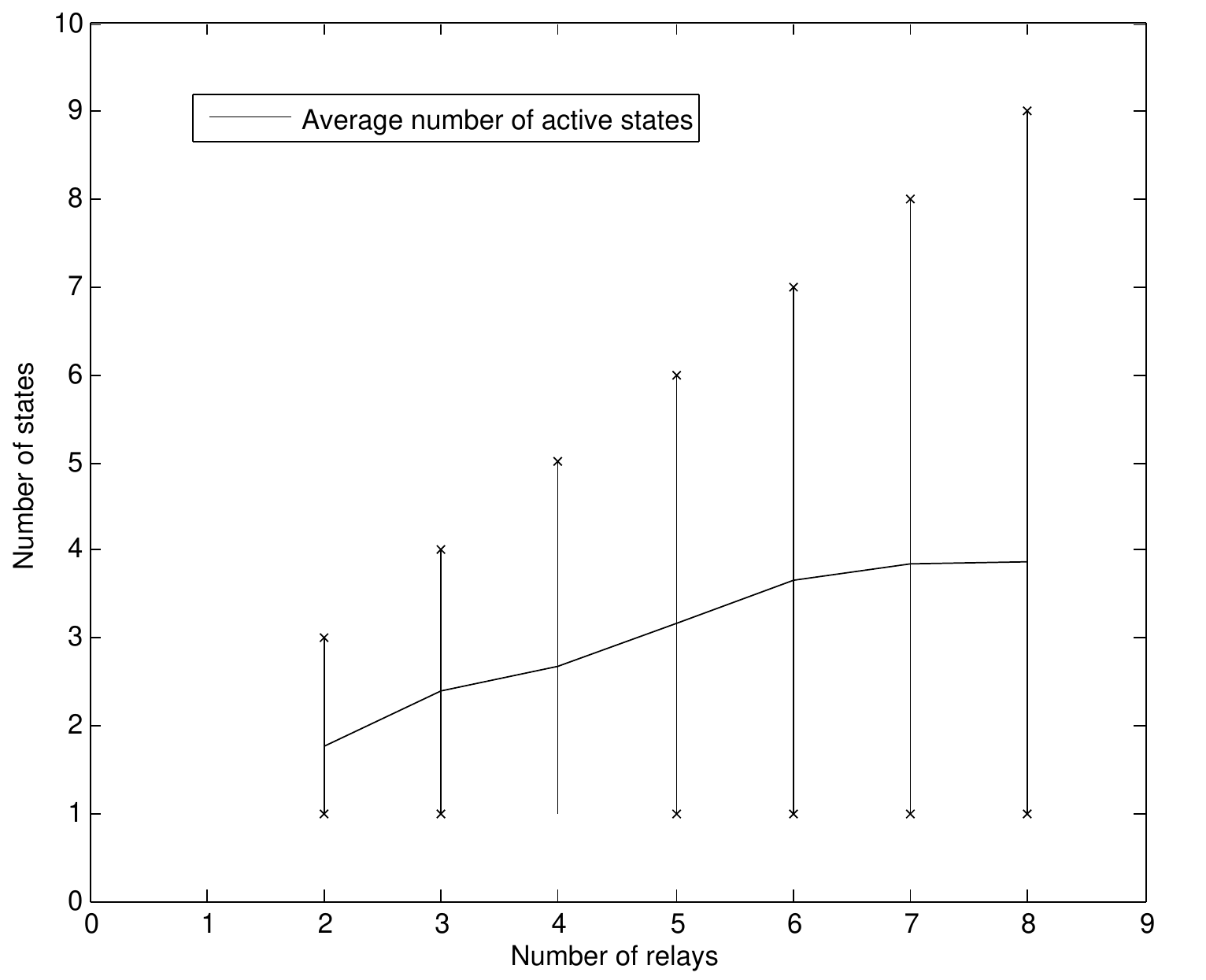}
\caption{Average, minimum and maximum number of active states to characterize the capacity of a HD relay network.}
\label{fig:activestates}
\end{figure*}

The conjecture holds for the case of $2$ relays as proved in Theorem \ref{thm:active2relays}. 
%For $K\!-\!2\!>\!2$ we have not been able to extend the proof of Thm. \ref{thm:active2relays}.
We proceeded through the following numerical evaluations: for each value of $N\leq 8$, we generated uniformly at random the SNR exponents of the channel gains, we computed the entries of $\mathbf{A}$ in~\eqref{eq:LP} with the Hungarian algorithm, we solved the LP in~\eqref{eq:LP} with the simplex method and we counted the number of constraints that equal the optimal gDoF (which is a known upper bound on the number of non zero entries of an optimal solution). The minimum and the maximum number of active states were found to be $1$ and $N+1$, respectively, as shown in Fig. \ref{fig:activestates}, which also shows the average number of active states computed by giving an equal weight to all the tried channels.
Note that the minimum number of active states for a {\em generic} HD relay network with $N$ relays has to be at least  $N+1$. To see this, consider a `line network' where the source can only communicate with relay~1, relay~1 can only communicate with relay~2, etc, and relay~$N$ can only communicate with the destination; in a line network, $N+1$ non-zero states are necessary to enable the source to communicate with the destination. It is interesting that the minimum number of active states given by $N+1$ also appears to be the required maximum number of active states for achieving the optimal gDoF-wise network operation.
If the reduction of the number of active states from exponential to linear as conjectured holds, it offers a simpler and more amenable way to design the network~\cite{Fragouli2012}.

\section{Fully-connected relay network with $ N=2$ relays}
\label{sec:example}
To gain insights into how relays are best utilized, we consider a network with $N=2$ relays. The analysis presented here differs from the one in~\cite{Fragouli2012} in the following:
(i) we study the fully-connected network, while in~\cite{Fragouli2012} only the diamond network is treated;
(ii) we explicitly find under which channel conditions the gDoF performance is enhanced by exploiting both relays instead of using only the best one, and
(iii) we provide insights into the nature of the rate gain in networks with multiple HD relays.
%and (iv) we compute the loss that HD incurs compared to FD. 

We consider the parameterization in \eqref{eq:par2rel} where, in order to increase the readability of the document, the SNR-exponents are indicated as 
%with $\beta_0=1$, i.e, the direct link from the source to the destination has gain $\mathsf{SNR}$ and all other channel gains are expressed with reference to it.
%\subsection{2-relay network}
%\label{sec:2relays}
%Here we analyze a relay network with $2$ relays, by parameterizing the channel gains as
\begin{align}
\left[\frac{\log(|h_{ij}|^2)}{\log(\mathsf{SNR})}\right]_{(i,j)\in[1:3]^2}
=
\begin{pmatrix}
\star   & \beta_1 & \alpha_{s1} \\ 
\beta_2 & \star   & \alpha_{s2} \\ 
\alpha_{1d} & \alpha_{2d} & 1 \\ 
\end{pmatrix},
\label{eq:par2rel}
\end{align}
where
$\star$ denotes an entry that does not matter for channel capacity (because a relay can remove the `self-interference'),
$\alpha_{si}$ is the SNR-exponent on the link from the source to relay~$i$, $i\in[1:2]$,
$\alpha_{id}$ is the SNR-exponent on the link from relay~$i$, $i\in[1:2]$, to the destination,
$\beta_i$ is the SNR-exponent on the link from relay~$j$ to relay~$i$, $(i,j)\in[1:2]^2$ with $j\not= i$, and
the direct link from the source to the destination (entry in position (3,3) in~\eqref{eq:par2rel}) has SNR-exponent normalized to 1 without loss of generality. Note that in order to consider a network without a direct link it suffices to consider all the other SNR-exponents to be larger than 1, or simply replace `1' with `0' in the discussion in the rest of the section.

We next derive the gDoF in both the FD and HD cases.

\subsection{The Full-Duplex Case}

For the FD case, the cut-set bound is achievable to within $2 \times 0.5105 \times 4 = 4.084$ bits with NNC (see Remark \ref{rem:remark3}).
As a consequence, it can be verified that the gDoF for the FD case is 
\begin{align}
\mathsf{d}_{N=2}^{\rm(FD)} 
  =
%  \lim_{\mathsf{SNR}\to+\infty} \frac{r^{\rm(CS-FD)}}{\log(1+\mathsf{SNR})} \nonumber\\&=
\min \Big\{
&\max \left\{ 1,\alpha_{s1},\alpha_{s2} \right \},
\max \left\{ \alpha_{s2}+\alpha_{1d},\beta_2+1 \right \}, \nonumber
\\& %\quad \quad \quad
\max \left\{ \alpha_{s1}+\alpha_{2d},\beta_1+1 \right \},
\max \left\{ 1,\alpha_{1d},\alpha_{2d} \right \} 
\Big \} \geq 1.
\label{eq:dofFD}
\end{align}
Notice that the gDoF in~\eqref{eq:dofFD} is no smaller than the gDoF that could be achieved by not using the relays, that is, by communicating directly through the direct link to achieve gDoF~$=1$. Notice also that the gDoF in~\eqref{eq:dofFD} does not change if we exchange $\alpha_{s1}$ with $\alpha_{2d}$ and $\alpha_{s2}$ with $\alpha_{1d}$, i.e., if we swap the role of the source and destination. 
We aim to identify the channel conditions under which using both relays strictly improves the gDoF compared to the best-relay selection policy (which includes direct transmission from the source to the destination as a special case) that achieves
%For future reference, if only one relay helps the communication between the source and the destination, and we choose the best among the two relays, then the achievable gDoF is \cite{AvestimehrThesis}
\begin{align}
\mathsf{d}_{N=2,\text{best relay}}^{\rm(FD)}
  &=
\max\big\{ 
1, \min\{\alpha_{s1},\alpha_{1d}\},  \min\{\alpha_{s2},\alpha_{2d}\}
\big\} \in [1, \mathsf{d}_{N=2}^{\rm(FD)}].
\label{eq:dofFDbestrealy}
\end{align}
%Note that, from the expression of $\mathsf{d}_{N=2}^{\rm(FD)}$ in~\eqref{eq:dofFD}, we immediately see that in order to have $\mathsf{d}_{N=2}^{\rm(FD)} > 1$, i.e., strictly better than direct transmission with the two relays silent,  we need
%\begin{align*}
%\max \left\{\alpha_{s1},\alpha_{s2} \right \} > 1, \quad
%\max \left\{\alpha_{s2}+\alpha_{1d}-1,\beta_2 \right \}>0, \\ 
%\max \left\{\alpha_{s1}+\alpha_{2d}-1,\beta_1 \right \}>0, \quad
%\max \left\{\alpha_{1d},\alpha_{2d} \right \} > 1.
%\end{align*}
We distinguish the following cases:

{\em Case~1):} 
if 
\[
\text{either} \ 
\left\{\begin{array}{l}
\alpha_{s1}\geq \alpha_{s2}\\
\alpha_{1d}\geq \alpha_{2d}\\
\end{array}\right.
\text{or} \ 
\left\{\begin{array}{l}
\alpha_{s1}< \alpha_{s2}\\
\alpha_{1d}< \alpha_{2d}\\
\end{array}\right.
\]
then, since one of the relays is `uniformly better' than the other, we immediately see that $\mathsf{d}_{N=2}^{\rm(FD)}= \mathsf{d}_{N=2,\text{best relay}}^{\rm(FD)}$, so in this regime selecting the best relay for transmission is gDoF optimal.

{\em Case~2):} 
if not in Case~1, then we are  in 
\[
\text{either} \ 
\left\{\begin{array}{l}
\alpha_{s1}\geq \alpha_{s2}\\
\alpha_{1d}< \alpha_{2d}\\
\end{array}\right.
\text{or} \ 
\left\{\begin{array}{l}
\alpha_{s1}< \alpha_{s2}\\
\alpha_{1d}\geq \alpha_{2d}\\
\end{array}\right..
\]
Consider the case $\alpha_{s2}\leq \alpha_{s1}, \  \alpha_{1d}< \alpha_{2d}$ (the other one is obtained essentially by swapping the role of the relays). This corresponds to an `asymmetric' situation where relay~1 has the best link from the source but relay~2 has the best link to the destination. In this case we would like to exploit the inter relay communication links (which is not present in a diamond network) to create a route source$\to$relay1$\to$relay2$\to$destination in addition to the direct link source$\to$destination.  Indeed, in this case $\mathsf{d}_{N=2}^{\rm(FD)}$ in~\eqref{eq:dofFD} can be rewritten as
\begin{align}
\mathsf{d}_{N=2}^{\rm(FD)} 
= 
\min \Big\{
\max \left \{ \alpha_{s2}+\alpha_{1d},\beta_2+1 \right \},
\max\{1,\min\{\alpha_{s1},\alpha_{2d}\} \}
\Big \},
\label{eq:dofFD case2}
\end{align}
where the term $\max\{1,\min\{\alpha_{s1},\alpha_{2d}\}\}$ in~\eqref{eq:dofFD case2} corresponds to the gDoF of a virtual single-relay channel such that the link from the source to the ``virtual relay'' has SNR-exponent $\alpha_{s1}$ and the link from the ``virtual relay'' to the destination has SNR-exponent $\alpha_{2d}$. 
%
%In order to do better than direct link transmission we must have $\min\{\alpha_{s1},\alpha_{2d}\} > 1$  {\blue and $\max \left\{\alpha_{s2}+\alpha_{1d}-1,\beta_2 \right \}>0$}.  {\red note that $1 \leq 1 +\beta_2 \leq \max \left \{ \alpha_{s2}+\alpha_{1d},\beta_2+1 \right \}$ so the only case we do as direct link transmission is if $\beta_2=0$ and $\alpha_{s2}+\alpha_{1d}\leq 1$}
%
%At this point we are left with the problem 
%\begin{align}
%\left\{\begin{array}{l}
%%\alpha_{s1} > 1 \\
%%\alpha_{2d} > 1 \\
%\min\{\alpha_{s1},\alpha_{2d}\} > 1 \\
%{\blue \max \left\{\alpha_{s2}+\alpha_{1d}-1,\beta_2 \right \}>0}\\
%\alpha_{s1}\geq \alpha_{s2}\\
%\alpha_{1d}< \alpha_{2d}\\
%\end{array}\right.
%: \
%\begin{array}{ll}
%&\mathsf{d}_{N=2}^{\rm(FD)} 
%= 
%\min \Big\{
%\max \left \{ \alpha_{s2}+\alpha_{1d},\beta_2+1 \right \},
%\alpha_{s1},\alpha_{2d} 
%\Big \} \\
%&> 
%\max\big\{ 
%1, \min\{\alpha_{s1},\alpha_{1d}\},  \min\{\alpha_{s2},\alpha_{2d}\}
%\big\}  = \mathsf{d}_{N=2,\text{best relay}}^{\rm(FD)} \\
%\end{array}
%\label{eq:were we might do better than relay selection}
%\end{align}
%We distinguish the following cases based on~\eqref{eq:were we might do better than relay selection}:
We aim to determine the subset of the channel parameters $\alpha_{s2}\leq \alpha_{s1}, \  \alpha_{1d}< \alpha_{2d}$ for which the gDoF in~\eqref{eq:dofFD case2} is strictly larger than the `best relay' gDoF in~\eqref{eq:dofFDbestrealy}.
The case
$
\alpha_{s2}\leq \alpha_{s1}, \ 
\alpha_{1d} <   \alpha_{2d}
%\label{eq:case 2 starting point}
$
subsumes the following possible orders of the channel gains:
\[
\begin{array}{|l| l l l l l l l l |}
\hline
\text{case i  }& \alpha_{1d} & \alpha_{2d} & \alpha_{s2} &             &             & \alpha_{s1} &             &             \\% & d_{BR}=(2b) & ? \\
\hline
\text{case ii }& \alpha_{1d} &             & \alpha_{s2} & \alpha_{2d} &             & \alpha_{s1} &             &             \\% & d_{BR}=(2a) & ? \\
\hline
\text{case iii}& \alpha_{1d} &             & \alpha_{s2} &             &             & \alpha_{s1} & \alpha_{2d} &             \\% & d_{BR}=(2a) & ? \\
\hline
\text{case iv }&             &             & \alpha_{s2} & \alpha_{1d} & \alpha_{2d} & \alpha_{s1} &             &             \\% & d_{BR}=(2a) & ? \\
\hline
\text{case v  }&             &             & \alpha_{s2} & \alpha_{1d} &             & \alpha_{s1} & \alpha_{2d} &             \\% & d_{BR}=(2a) & ? \\
\hline
\text{case vi }&             &             & \alpha_{s2} &             &             & \alpha_{s1} & \alpha_{1d} & \alpha_{2d} \\% & d_{BR}=(2c) & ? \\
\hline
\end{array}
\]
We partition the set of channel parameters $\alpha_{s2}\leq \alpha_{s1}, \  \alpha_{1d}< \alpha_{2d}$ as follows:
\begin{itemize}

\item {\em Sub-case~2a)} 
%{\magenta
(all but cases i and vi in the table above):
%}
if
\begin{align}
\max \{\alpha_{s2},\alpha_{1d}\}  < \min\{\alpha_{s1},\alpha_{2d}\}, \ \
\label{eq:cond2a} 
\end{align}
then
\begin{align}
\mathsf{d}_{N=2,\text{best relay}}^{\rm(FD)} 
= \max \{1, \alpha_{s2},\alpha_{1d}\}, 
%= \max \Big\{1, \max \{\alpha_{s2},\alpha_{1d}\}  \Big\},
\label{eq:sub-case 2a dbestrelay}
\end{align}
which is strictly less than  $\mathsf{d}_{N=2}^{\rm(FD)}$ in \eqref{eq:dofFD case2} if
%\begin{subequations}
\begin{align*}
&\text{ either} 
&&\max \{1, \alpha_{s2},\alpha_{1d}\}  < \min\{\alpha_{s1},\alpha_{2d}\}\leq \max \left \{\alpha_{s2}+\alpha_{1d},\beta_2+1\right \}
\\&\text{ or} 
&&\Big\{ \max \left \{\alpha_{s2}+\alpha_{1d},\beta_2+1\right \} < \min\{\alpha_{s1},\alpha_{2d}\}\Big\} \cap \mathcal{O}^c
%\end{align*}
%\end{subequations}
%where
%\begin{align*}
\\&\text{ where} 
&&\mathcal{O} \!:=\! 
\{ \beta_2\!=\!0, \alpha_{s2}\!+\!\alpha_{1d}\!\leq\! 1\}\cup
\{ \alpha_{1d}\!=\!0, \beta_2\!+\!1\!\leq\! \alpha_{s2}\}\cup
\{ \alpha_{s2}\!=\!0, \beta_2\!+\!1\!\leq\! \alpha_{1d}\}.
\end{align*}
that is, for
% {\red RIGHT?}
\begin{align}
\max \{1, \alpha_{s2},\alpha_{1d}\}  < \min\{\alpha_{s1},\alpha_{2d}\} \ \text{except in region $\mathcal{O}$}.
\label{eq:sub-case 2a finalconditionatlast}
\end{align}

\item {\em Sub-case~2b)} (case i in the table above):
% when not in ub-case~2a we have
%1 < \min\{\alpha_{s1},\alpha_{2d}\} \leq  \max \{\alpha_{s2},\alpha_{1d}\} 
if
%\begin{align*}
%\{ \beta_2=0, \alpha_{s2}+\alpha_{1d}\leq 1 \}^c, \ \ 
%1 < \alpha_{2d}, \ \ 
$\alpha_{1d} < \alpha_{2d} \leq \alpha_{s2} \leq \alpha_{s1}$, 
%\end{align*}
then the condition
\[
\mathsf{d}_{N=2,\text{best relay}}^{\rm(FD)} = \max\{1,\alpha_{2d}\}
< %\stackrel{?}{<} 
\mathsf{d}_{N=2}^{\rm(FD)} = \min \Big\{
\max \left \{ \alpha_{s2}+\alpha_{1d},\beta_2+1 \right \},
\max\{1,\alpha_{2d} \}
\Big \}
\]
is never verified, i.e., in this case $\mathsf{d}_{N=2,\text{best relay}}^{\rm(FD)}=\mathsf{d}_{N=2}^{\rm(FD)}$.

\item {\em Sub-case~2c)} (case vi in the table above):
if
%\begin{align*}
%\{ \beta_2=0, \alpha_{s2}+\alpha_{1d}\leq 1 \}^c, \ \   
%1 < \alpha_{s1}, \ \ 
$\alpha_{s2} \leq \alpha_{s1} \leq \alpha_{1d} < \alpha_{2d}$,
%\end{align*}
then
\[
\mathsf{d}_{N=2,\text{best relay}}^{\rm(FD)} = \max\{1,\alpha_{s1}\}
< %\stackrel{?}{<} 
\mathsf{d}_{N=2}^{\rm(FD)} = \min \Big\{
\max \left \{ \alpha_{s2}+\alpha_{1d},\beta_2+1 \right \},
\max\{1,\alpha_{s1}\}
\Big \}
\]
is never verified, i.e., in this case $\mathsf{d}_{N=2,\text{best relay}}^{\rm(FD)}=\mathsf{d}_{N=2}^{\rm(FD)}$.

\end{itemize}
Recall that there is also a regime similar Case~2) where the role of the relays is swapped.

\begin{figure*}
\centering
\includegraphics[width=0.7\textwidth]{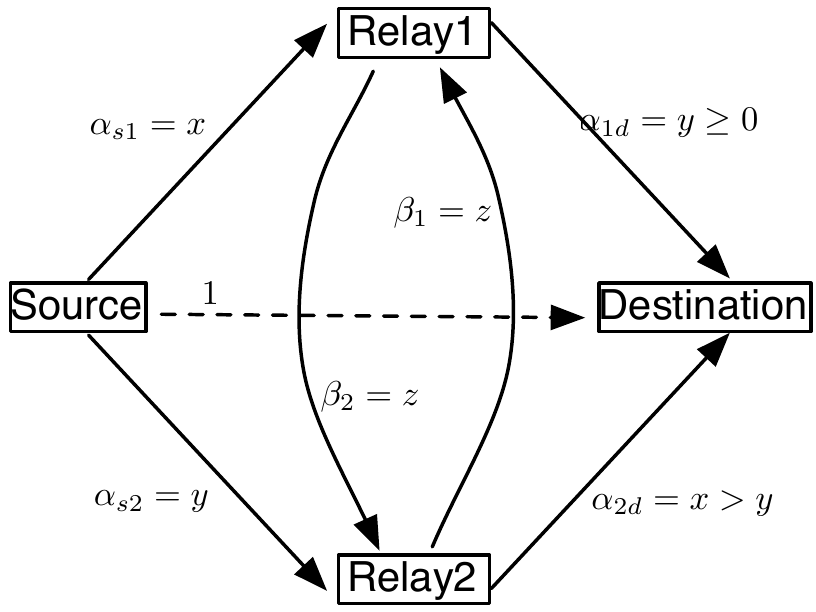}
\caption{Example of a two-relay network.}
% with gDoF strictly larger than the gDoF obtained by using the best relay only.}
\label{fig:tworelaynet}
\end{figure*}

{
Fig.~\ref{fig:tworelaynet} gives an example of a network satisfying the conditions in~\eqref{eq:cond2a}, i.e., the assumption is $0\leq y < x$ without loss of generality.
%for which, by using both relays, the system achieves a strictly greater gDoF compared to the one that would be attained by exploiting only the best relay by keeping the other silent.  The numerical value on a link represents the SNR exponent on the corresponding link.
%
By exploiting both relays, the system attains
\begin{align*}
\mathsf{d}_{N=2}^{\rm(FD)} 
&=
\min \Big\{
\max\{ 1,x,y \},
\max\{ 2x,z+1\},
\max\{ 2y,z+1\}
\Big \}
\\&=
\min \Big\{
\max\{ 1,x   \},
\max\{ 2y,z+1 \}
\Big \},
\end{align*}
while, by using only the best relay, it achieves
\begin{align*}
\mathsf{d}_{N=2,\text{best relay}}^{\rm(FD)}
&=
\max\big\{ 
1, \min\{x,y\}
\big\}=\max\big\{1,y\big\}.
\end{align*}
By~\eqref{eq:sub-case 2a finalconditionatlast}, we have $\mathsf{d}_{N=2}^{\rm(FD)} > \mathsf{d}_{N=2,\text{best relay}}^{\rm(FD)}$  if 
\begin{align}
x>\max\big\{1,y\big\} \ \text{except} \ \left\{ z=0, y\leq \frac{1}{2} \right\}.
\label{eq:sub-case 2a finalconditionatlast example}
\end{align}
}

\subsection{The Half-Duplex Case}
%With HD, each term in the min-function defining the cut-set bound is a convex combination of $2^{N}$ terms, one for each possible state of the network. Formally, these terms are as for the FD case but where one replaces the SNR exponent of the link from node~$j$ to node~$i$, given by $\alpha_{ij}$, with $(1-S_i) \alpha_{ij} S_j$, where $(S_i,S_j)\in\{0,1\}^2$ indicates the state of the nodes. Because no additional insight is gained from these terms, we do not report them here. At high SNR the outer bound gives the following gDoF
With HD, the gDoF is given by~\eqref{eq:LP}, which with the notation in~\eqref{eq:par2rel} and with $\lambda_1=\lambda_{00}$, $\lambda_2=\lambda_{01}$, $\lambda_3=\lambda_{10}$ and $\lambda_4=\lambda_{11}$ becomes
\begin{align}
\mathsf{d}_{N=2}^{\rm(HD)} 
=  \max
\min\Big\{
  & \lambda_{00} D_1^{(0)}+\lambda_{01} D_1^{(1)}+\lambda_{10} D_1^{(2)}+\lambda_{11} D_1^{(3)}, \nonumber
\\& \lambda_{00} D_2^{(0)}+\lambda_{01} D_2^{(1)}+\lambda_{10} D_2^{(2)}+\lambda_{11} D_2^{(3)}, \nonumber
\\& \lambda_{00} D_3^{(0)}+\lambda_{01} D_3^{(1)}+\lambda_{10} D_3^{(2)}+\lambda_{11} D_3^{(3)}, \nonumber
\\& \lambda_{00} D_4^{(0)}+\lambda_{01} D_4^{(1)}+\lambda_{10} D_4^{(2)}+\lambda_{11} D_4^{(3)}
\Big\},
\label{eq:dofHD}
\end{align}
where the maximization is over $\lambda_{ij}=\mathbb{P}[S_1=i,S_2=j] \geq 0$, $(i,j) \in \{ 0,1\}^2$, such that 
$\lambda_{00}+\lambda_{01}+\lambda_{10}+\lambda_{11}=1$ and
\begin{align*}
\begin{array}{ll}
   D_1^{(0)} := \max \left \{ 1,\alpha_{s1},\alpha_{s2} \right \},
 & D_1^{(1)} = D_2^{(0)} := \max \left \{ 1,\alpha_{s1} \right \},
\\ D_4^{(3)} := \max \left \{ 1,\alpha_{1d},\alpha_{2d} \right \},
 & D_1^{(2)} = D_3^{(0)} := \max \left \{ 1,\alpha_{s2} \right \},  
\\ D_2^{(1)} :=  \max  \left \{  \alpha_{s1}+\alpha_{2d}, \beta_1 +1 \right \},  
 & D_2^{(3)} = D_4^{(1)} := \max \left \{ 1,\alpha_{2d} \right \}, 
\\ D_3^{(2)} := \max \left \{ \alpha_{s2}+\alpha_{1d},\beta_2 +1 \right \},
 & D_3^{(3)} = D_4^{(2)} :=\max \left \{ 1,\alpha_{1d} \right \},
\\ D_1^{(3)} = D_2^{(2)} = D_3^{(1)} = D_4^{(0)} := 1.  
\end{array}
\end{align*}
For future reference, if only one relay helps the communication between the source and the destination then the achievable gDoF is \cite{cardoneISIT2013}
\begin{align}
\mathsf{d}_{N=2,\text{best relay}}^{\rm(HD)}
  &= 1+ \max_{i\in[1:2]}\frac{[\alpha_{si}-1]^+ \ [\alpha_{id}-1]^+}{[\alpha_{si}-1]^+ + [\alpha_{id}-1]^+}
 \in [1, \mathsf{d}_{N=2}^{\rm(HD)}].
\label{eq:dofHDbestrealy}
\end{align}

An analytical closed form solution for the optimal $\{\lambda_{ij}\}$ in~\eqref{eq:dofHD} is complex to find for general channel gain assignments.  However, numerically it is a question of solving a LP, for which efficient numerical routines exist. By using Theorem~\ref{thm:active2relays}, we can set either $\lambda_{00}$ or $\lambda_{11}$ to zero. 
%We remark that this LP can be thought as the high SNR solution of the iterative algorithm proposed in~\cite{OngMultiRelay}.
%
%For the example presented in Fig.~\ref{fig:tworelaynet}, it is possible to find the gDoF $\mathsf{d}_{N=2}^{\rm(HD)}$ in closed form. 

For the example in Fig.~\ref{fig:tworelaynet}  the optimal schedule has $\lambda_{00}=\lambda_{11}=0$ without loss of optimality, from Theorem \ref{thm:active2relays}. 
%BECAUSE $[X-1]^+ [Y-1]^+ \leq [X-1]^+ [Y-1]^+$
By letting $\lambda_{01}=\gamma\in[0,1]$ and $\lambda_{10}=1-\gamma$ (recall $0\leq y < x$ without loss of generality), the gDoF in \eqref{eq:dofHD} can be written as
\begin{subequations}
\begin{align}
\mathsf{d}_{N=2}^{\rm(HD)} 
=  \max_{\gamma\in[0,1]}
\min\Big\{
  & \gamma \max \{1,x\} + (1-\gamma) \max \{1,y\}, \label{eq:example1}
\\& \gamma \max \{2x,z+1 \}+(1-\gamma),\label{eq:example2}
\\&\gamma+(1-\gamma) \max \{2y,z+1 \}\label{eq:example3}
\Big\}
\\= 1+\min &\left \{\frac{[x-1]^+  \max \{2y-1,z \}}{[x-1]^++\max \{2y-1,z \}-[y-1]^+},\right. 
\\&
\left. \frac{\max \{2x-1,z \}\max \{2y-1,z \}}{\max \{2x-1,z \}+\max \{2y-1,z \}}\right \}.
\end{align}
\end{subequations}
%{\blue
%The terms \eqref{eq:example1} and \eqref{eq:example2} are always increasing in $\gamma$,
%while the term \eqref{eq:example3} is always decreasing in $\gamma$. Therefore, the two optimal $\gamma$ are obtained by equating \eqref{eq:example1} with \eqref{eq:example3} and \eqref{eq:example2} with \eqref{eq:example3}. These two values of $\gamma$, then lead to two different values of gDoF; the attained gDoF is the minimum among these two possible values, that is \eqref{eq:DoFCF} holds.
%}
By using only the best relay as in \eqref{eq:dofHDbestrealy}, we would achieve
\begin{align}
\mathsf{d}_{N=2,\text{best relay}}^{\rm(HD)}
  &= 1+\frac{[x-1]^+ [y-1]^+}{[x-1]^++[y-1]^+}.
\label{eq:bestrelayexsym}
\end{align}
It can be easily seen that the best relay selection policy is strictly suboptimal if~\eqref{eq:sub-case 2a finalconditionatlast example} is verified, as for the FD case.

% DT I DO NOT THIK WE NEED THIS 
%Table~\ref{table:nonlin} shows the gDoF corresponding to the four cases where using both relays strictly improves over exploiting only the best relay. 
%%We denote with $\mathsf{d}_{N=2,\text{best relay}}^{\rm(HD)}$ the gDoF that is obtained when the two relays work in HD by selecting the relay which achieves the highest gDoF while leaving the other silent, whose closed form solution is given in~\cite{cardoneISIT2013}.
%From Table \ref{table:nonlin} we notice that in each case where  $\mathsf{d}_{N=2}^{\rm(FD)}>\mathsf{d}_{N=2}^{\rm(HD)}$ also $\mathsf{d}_{N=2}^{\rm(HD)} > \mathsf{d}_{N=2,\text{best relay}}^{\rm(HD)}$.
%\begin{table}[!h]
%\centering
%\caption{gDoF when using both relays is better than using only the best one.} 
%\label{table:nonlin} 
%\begin{tabular}{c c c c c} 
%\toprule 
%\textbf{Channel Parameters} & \textbf{Full-duplex} & \textbf{Full-duplex} & \textbf{Half-duplex} & \textbf{Half-duplex}\\
%\textbf{$\left( \alpha_{s1},\alpha_{s2},\alpha_{1d},\alpha_{2d},\beta_1,\beta_2 \right)$} & \textbf{best relay} & \textbf{both relays} & \textbf{best relay} & \textbf{both relays} \\
%\midrule
%$\left( 2.5,1.4,0.5,1.8,0.6,0.8 \right)$ & $1.4$ & $1.8$ & $1.267$ & $1.4235$ \\ 
%$\left( 2.5,0.3,0.7,1.3,0.4,0.8 \right)$ & $1.0$ & $1.3$ & $1.000$ & $1.2182$ \\
%$\left( 1.8,1.2,1.3,2.0,0.7,1.2 \right)$ & $1.3$ & $1.8$ & $1.218$ & $1.5808$ \\
%$\left( 1.7,1.1,1.2,1.4,0.4,1.5 \right)$ & $1.2$ & $1.4$ & $1.156$ & $1.3604$ \\
%\bottomrule
%\end{tabular}
%\end{table}

\section{Conclusions}
\label{sec:conl}
In this work we analyzed a network where a source communicates with a destination across a Gaussian channel. This communication is assisted by $N$ relays operating in half-duplex mode. We characterized the capacity to within a constant gap by using an achievable scheme based on noisy network coding. We also showed that this gap may be further reduced by considering more structured systems, such as the diamond network. We conjectured that the optimal schedule has at most $N+1$ active states, instead of the possible $2^{N}$. This conjecture has been supported by the analytical proof in the special case of $N=2$ relays and in general by numerical evaluations. 
We finally analyzed a network with $N=2$ relays, and we showed under which channel conditions by exploiting both relays a strictly greater gDoF can be attained compared to a network where best-relay selection is used.

An interesting connection between the high-SNR approximation of the point-to-point MIMO capacity and the Maximum Weighted Bipartite Matching problem from graph theory has been discovered.

%\newpage
\appendices

\section{Proof of Proposition~\ref{prop:hd diamond}}
\label{app:hd diamond}
In a multicast network, where $\mathbf{H}_{\rm r \to r}=0$ and $\mathbf{H}_{\rm s \to d}=0$ in~\eqref{eq:channel2}, the rank of any channel sub matrix is upper bounded by $2$. Thus, with ${\rm Rank}[\mathbf{H}_{\mathcal{A},s}] \leq 2$ in the cut-set bound in the step preceding \eqref{eq:eqnewmulti} and by using the NNC lower bound in \eqref{eq:eqnewmultilow}, the gap becomes
\begin{align*}
\mathsf{GAP} & \leq \min_{\gamma \geq \eu-1} \max_{\mu\in[0:1]} \left \{2 \min \left \{  \frac{\gamma}{\eu},\frac{2}{\mu K}\right \} \log \left( \max \left \{ \eu,\frac{\gamma \ \mu K}{2} \right \} \right)\right.
 \left.+\mu K \log\left(\frac{2\gamma}{\gamma-1}\right)\right \}
\\& \stackrel{(a)}{=} \min_{\gamma \geq \eu-1} \left. \left \{ 2 \min \left \{  \frac{\gamma}{\eu},\frac{2}{K}\right \} \log \left(\eu \max \left \{ 1,\frac{\gamma}{\eu} \frac{K}{2} \right \} \right) \right. \right.
\left. + K \log\left(\frac{2\gamma}{\gamma-1}\right) \right \} %\left. \right |_{x:=\frac{\gamma}{\eu}}
\\& \stackrel{(b)}{=} \left \{
\begin{array}{ll}
2 \log (\eu) \frac{1+\sqrt{1+2 K \eu}}{2 \eu}    + K \log \left( \frac{2\sqrt{1+2 K \eu}+2}{\sqrt{1+2 K \eu}-1} \right) & \text{if}\ K \leq 2
\\\frac{4}{K} \log \left( \frac{K}{2}+\frac{K^3}{8}\right)+K\log \left( 2+\frac{8}{K^2} \right) & \text{if} \ K>2
\end{array},
\right.
\end{align*}
where: the equality in (a) follows since the function is always increasing in $\mu$ so the maximum is attained for $\mu^{\rm{opt}}=1$; the minimum over $\gamma$ is attained for 
\begin{align*}
\gamma^{\rm opt}=\left \{
\begin{array}{ll}
\frac{1+\sqrt{1+2 K \eu}}{2} & \text{if} \ K\leq2
\\ 1+\frac{K^2}{4} & \text{if} \ K>2
\end{array}
\right.
\end{align*}
and this leads to the equality in (b). 

By substituting $K=N+2$ in order to obtain the special case of the HD multi-relay diamond network we get \eqref{eq:gap diamond}.

\section{Proof of Theorem~\ref{th:MWBM}}
\label{app:MWBM}

Let $\mathcal{S}_{n,k}$ be the set of all $k$-combinations of the integers in $[1:n]$ and $\mathcal{P}_{n,k}$ be the set of all $k$-permutations of the integers in $[1:n]$. Let $\sigma(\pi)$ be the sign / signature of the permutation $\pi$.

We start by demonstrating that the asymptotic behavior of $|\mathbf{I}_k+\mathbf{H}\mathbf{H}^H|$ is as that of $|\mathbf{H}\mathbf{H}^H|$, i.e., the identity matrix can be neglected. By using the determinant Leibniz formula \cite{Broida}, in fact we have, 
\begin{align*}
&|\mathbf{I}_k+\mathbf{H}\mathbf{H}^H|  
= \sum_{\pi\in \mathcal{P}_{n,k}} \sigma(\pi) \prod_{i=1}^{k} \left[ \mathbf{I}_k+\mathbf{H}\mathbf{H}^H \right ]_{i,\pi(i)}
\\& = \sum_{\pi \in \mathcal{P}_{n,k}} \sigma(\pi) \left \{ \left ( \prod_{i=2}^{k}  \left[ \mathbf{I}_k+\mathbf{H}\mathbf{H}^H \right ]_{i,\pi(i)} \right ) \left ( \left[ \mathbf{I}_k+\mathbf{H}\mathbf{H}^H \right ]_{1,\pi(1)} \right ) \right \}
\\& = \sum_{\pi \in \mathcal{P}_{n,k}} \sigma(\pi) \left ( \prod_{i=2}^{k}  \left[ \mathbf{I}_k+\mathbf{H}\mathbf{H}^H \right ]_{i,\pi(i)} \right ) \delta[1-\pi(1)]
\\& + \sum_{\pi\in \mathcal{P}_{n,k}} \sigma(\pi)\prod_{i=2}^{k} \left[ \mathbf{I}_k+\mathbf{H}\mathbf{H}^H \right ]_{i,\pi(i)} \left[ \mathbf{H}\mathbf{H}^H \right ]_{1,\pi(1)},
\end{align*}
where $\delta[n] = \left \{ \begin{array}{ll} 0 & n \neq 0 \\ 1 & n=0 \end{array}\right.$ is the Kronecker delta.
Let
\begin{align*}
  &\mathsf{A} \left( \mathsf{SNR}\right):=  \sum_{\pi\in \mathcal{P}_{n,k}} \sigma(\pi) \left ( \prod_{i=2}^{k}  \left[ \mathbf{I}_k+\mathbf{H}\mathbf{H}^H \right ]_{i,\pi(i)} \right ) \delta[1-\pi(1)]
\\& \mathsf{B} \left( \mathsf{SNR}\right):= \sum_{\pi\in \mathcal{P}_{n,k}} \sigma(\pi)\prod_{i=2}^{k} \left[ \mathbf{I}_k+\mathbf{H}\mathbf{H}^H \right ]_{i,\pi(i)} \left[ \mathbf{H}\mathbf{H}^H \right ]_{1,\pi(1)},
\end{align*}
we have that
$\mathsf{A} \left( \mathsf{SNR}\right) = o \left( \mathsf{B} \left( \mathsf{SNR}\right)\right ), \ \text{because} \  \lim \limits_{\mathsf{SNR} \to +\infty} \frac{\mathsf{A} \left( \mathsf{SNR}\right)}{ \mathsf{B} \left( \mathsf{SNR}\right)}=0$ where the $\mathsf{SNR}$ parameterizes the channel gains as $|h_{ij}|^2=\mathsf{SNR}^{\beta_{ij}}$, for some non-negative $\beta_{ij}$. This is so because, as a function of $\mathsf{SNR}$, $\mathsf{B} \left( \mathsf{SNR}\right)$ grows faster than $\mathsf{A} \left( \mathsf{SNR}\right)$ due to the term $\left[ \mathbf{H}\mathbf{H}^H \right ]_{1,\pi(1)}$.
By induction  it is possible to show that this reasoning holds $\forall i \in[1:k]$ and hence
\begin{align*}
&|\mathbf{I}_k+\mathbf{H}\mathbf{H}^H|
\doteq 
 \sum_{\pi\in \mathcal{P}_{n,k}} \sigma(\pi) \prod_{i=1}^{k} \left[ \mathbf{H}\mathbf{H}^H \right ]_{i,\pi(i)}
=|\mathbf{H}\mathbf{H}^H|.
\end{align*}
% This shows that the asymptotic behavior of $|\mathbf{I}_k+\mathbf{H}\mathbf{H}^H|$ is as that of $|\mathbf{H}\mathbf{H}^H|$. 
Therefore, now we focus on the study of $|\mathbf{H}\mathbf{H}^H|$. We have
\begin{align*}
&|\mathbf{H}\mathbf{H}^H| 
\stackrel {(a)}{=} \sum_{\varsigma \in \mathcal{S}_{n,k}} |\mathbf{H}_\varsigma| |\mathbf{H}_\varsigma^H|
 = \sum_{\varsigma \in \mathcal{S}_{n,k}}  |\mathbf{H}_\varsigma|^2 
 \\&\stackrel {(b)}{=}  \sum_{\varsigma \in \mathcal{S}_{n,k}} \left |  \sum_{\pi \in \mathcal{P}_{n,k}} \sigma(\pi) \prod_{i=1}^{k} \left [ \mathbf{H}_\varsigma \right ]_{i,\pi(i)} \right |^2
\\& = \sum_{\varsigma \in \mathcal{S}_{n,k}}  \left \{  \left (\sum_{\pi_1 \in \mathcal{P}_{n,k}} \sigma(\pi_1) \prod_{i=1}^{k} \left [ \mathbf{H}_\varsigma \right ]_{i,{\pi_1}(i)} \right ) \left( \sum_{\pi_2 \in \mathcal{P}_{n,k}} \sigma(\pi_2) \prod_{j=1}^{k} \left [ \mathbf{H}_\varsigma \right ]_{j,{\pi_2}(j)} \right )^* \right \}
\\& = \sum_{\varsigma \in \mathcal{S}_{n,k}}  \left \{  \left (\sum_{\pi  \in \mathcal{P}_{n,k}} \prod_{i=1}^{k} \left | \left [ \mathbf{H}_\varsigma \right ]_{i,{\pi}(i)} \right |^2 \right )\right.
\\& \left.+\left( \sum_{\pi_1,\pi_2 \in \mathcal{P}_{n,k},\pi_1\neq \pi_2} \sigma (\pi_1) \sigma (\pi_2) \prod_{i=1}^{k} \prod_{j=1}^{k} \left [ \mathbf{H}_\varsigma \right ]_{i,{{\pi}_1}(i)}\left(\left [ \mathbf{H}_\varsigma \right ]_{j,{{\pi}_2}(j)}\right)^* \right)\right \}
\\& \stackrel {(c)}{\leq} \sum_{\varsigma \in \mathcal{S}_{n,k}}  \left \{  \left (\sum_{\pi  \in \mathcal{P}_{n,k}} \prod_{i=1}^{k} \left | \left [ \mathbf{H}_\varsigma \right ]_{i,{\pi}(i)} \right |^2 \right ) \right.
\\&\left.+ \left( \sum_{\pi_1,\pi_2 \in \mathcal{P}_{n,k},\pi_1\neq \pi_2}\prod_{i=1}^{k} \prod_{j=1}^{k} \sqrt{\left | \left [ \mathbf{H}_\varsigma \right ]_{i,{{\pi}_1}(i)} \right |^2\left | \left [ \mathbf{H}_\varsigma \right ]_{j,{{\pi}_2}(j)} \right |^2} \right) \right \}
\\& = \sum_{\varsigma \in \mathcal{S}_{n,k}}  \left \{  \left (\sum_{\pi  \in \mathcal{P}_{n,k}} \mathsf{SNR}^{\sum_{i=1}^k \left [ \mathbf{B}_\varsigma \right ]_{i,\pi(i)}}  \right ) \right.
\\&\left.+ \left( \sum_{\pi_1,\pi_2 \in \mathcal{P}_{n,k},\pi_1\neq \pi_2} \mathsf{SNR}^{\frac{1}{2}\left(\sum_{i=1}^k \left [ \mathbf{B}_\varsigma \right ]_{i,{\pi_1}(i)}+\sum_{j=1}^k \left [ \mathbf{B}_\varsigma \right ]_{j,{\pi_2}(j)}\right)} \right) \right \}
\\& \stackrel {(d)}{\doteq}\sum_{\varsigma \in \mathcal{S}_{n,k}} \left (\sum_{\pi  \in \mathcal{P}_{n,k}} \mathsf{SNR}^{\sum_{i=1}^k \left [ \mathbf{B}_\varsigma \right ]_{i,\pi(i)}}  \right ) 
%\\& \stackrel {\mathsf{SNR}\rightarrow \infty}{\doteq} \sum_{\varsigma \in \mathcal{S}_{n,k}} \mathsf{SNR}^{\max_{\pi  \in \mathcal{P}_{n,k}}\left [ \beta_\varsigma \right ]_{i,\pi(i)}}
\\& %\stackrel {\mathsf{SNR}\rightarrow \infty}
{\doteq} \ \mathsf{SNR}^{
\max_{\varsigma \in \mathcal{S}_{n,k}}
\max_{\pi  \in \mathcal{P}_{n,k}}
\sum_{i=1}^k 
\left [ \mathbf{B}_\varsigma \right ]_{i,\pi(i)}},
\end{align*}
where the equalities /  inequalities above are due to the following facts:
\begin{itemize}
\item equality (a): by applying the Cauchy-Binet formula \cite{Broida} where $\mathbf{H}_\varsigma$ is the square matrix obtained from $\mathbf{H}$ by retaining all rows and those columns indexed by $\varsigma$;
\item equality (b): by applying the determinant Leibniz formula \cite{Broida};
\item inequality (c): by applying the Cauchy-Swartz inequality \cite{Steele};
\item equality (d): when $\mathsf{SNR}\rightarrow \infty$, we have 
\begin{align*}
\sum_{i=1}^k \left [ \mathbf{B}_\varsigma \right ]_{i,\pi(i)} \geq \frac{1}{2}\left(\sum_{i=1}^k \left [ \mathbf{B}_\varsigma \right ]_{i,{\pi_1}(i)}+\sum_{j=1}^k \left [ \mathbf{B}_\varsigma \right ]_{j,{\pi_2}(j)}\right). 
\end{align*}
Consider the following example. Let $a=\mathsf{SNR}^{\beta_a}$, $b=\mathsf{SNR}^{\beta_b}$, $c=\mathsf{SNR}^{\beta_c}$, $d=\mathsf{SNR}^{\beta_d}$
\begin{align*}
|ab-cd|^2 \leq |a|^2|b|^2+|c|^2|d|^2+2|a||b||c||d|.
\end{align*}
Now apply the gDoF formula, i.e.,
\begin{align*}
\mathsf{d} &:= \lim \limits_{\mathsf{SNR} \to +\infty} \frac{\log \left(|a|^2|b|^2+|c|^2|d|^2+2|a||b||c||d|\right)}{\log(1+\mathsf{SNR})}
\\&= \max \left\{  \beta_a+\beta_b,\beta_c+\beta_d, \frac{\beta_a+\beta_b+\beta_c+\beta_d}{2}\right \},
\end{align*}
but
\begin{align*}
\frac{\beta_a+\beta_b+\beta_c+\beta_d}{2} \leq \frac{2\max \left\{ \beta_a+\beta_b,\beta_c+\beta_d\right\}}{2}=\max \left\{ \beta_a+\beta_b,\beta_c+\beta_d \right \}. 
\end{align*}
Therefore, the term $\frac{\beta_a+\beta_b+\beta_c+\beta_d}{2}$ does not contribute in characterizing the gDoF.
By direct induction, the above reasoning may be extended to a general number of terms leading to $\sum_{i=1}^k \left [ \mathbf{B}_\varsigma \right ]_{i,\pi(i)} \geq \frac{1}{2}\left(\sum_{i=1}^k \left [ \mathbf{B}_\varsigma \right ]_{i,{\pi_1}(i)}+\sum_{j=1}^k \left [ \mathbf{B}_\varsigma \right ]_{j,{\pi_2}(j)}\right)$.
\end{itemize}

%\section{Proof of Proposition~\ref{prop:hd reduced gap}}
%\label{app:redGap}
%The proof directly follows from the proof of Theorem \ref{thm:many relays}, with the following difference. Instead of the upper bound $H(S_{\mathcal{A}}) \leq |\mathcal{A}| \log(2)$ we proceed as follows
%\[
%H(S_{\mathcal{A}}) \leq H(S_{[2: N+2-2]}) \leq \log\left(\sum_{s=0}^{2^{N}-1} {\mathbf{1}}_{\{\lambda_s >0\}}\right)
%\]
%where ${\mathbf{1}}_{\{\lambda_s >0\}}$ is the indicator function defined as
%\begin{align*}
%{\mathbf{1}}_{\{\lambda_s >0\}} = \left \{ 
%\begin{array}{ll}
%1 & {\text{if}} \ \lambda_s >0\\
%0 & {\text{otherwise}}
%\end{array}.
% \right.
%\end{align*}
%If Conjecture \ref{conj} holds, we have that $\sum_{s=0}^{2^{N}-1} {\mathbf{1}}_{\{\lambda_s^\star >0\}} \leq  N+2-1$ and so our gap would be at most
%\begin{align*}
%\mathsf{GAP} &\leq \max_{|\mathcal{A}|\in[0, N]}\Big\{
%\min(1+|\mathcal{A}|, 1+|\mathcal{A}^c|)  \log\left(1+|\mathcal{A}|\right)
%+\min( 1+2|\mathcal{A}|, N+2-1)\log(2)
%\Big\}
%\\& +\log \left(  N+2-1\right ).
%\end{align*}

\section{Proof of Theorem~\ref{thm:active2relays}}
\label{app:hdactive2}
In a HD relay network with $N=2$, we have $4$ possible states that may arise with probabilities $\lambda_j$ with $j \in[1,4]$. We let $\lambda_1=\lambda_{00}$, $\lambda_2=\lambda_{01}$, $\lambda_3=\lambda_{10}$ and $\lambda_4=\lambda_{11}$, where $\lambda_{ij}=\mathbb{P}[S_1=i,S_2=j] \geq 0$, $(i,j) \in \{ 0,1\}^2$, such that 
$\lambda_{00}+\lambda_{01}+\lambda_{10}+\lambda_{11}=1$. Here we aim to demonstrate that a schedule with $\lambda_{00}\lambda_{11}=0$ is optimal, i.e., it maximizes the capacity of the HD relay network with $N=2$.
Let 
\begin{align*}
  &\alpha_{s1}-1 :=\alpha^\star_{s1}, \quad \alpha_{s2}-1 :=\alpha^\star_{s2},
\\&\alpha_{1d}-1 :=\alpha^\star_{1d}, \quad \alpha_{2d}-1 :=\alpha^\star_{2d}. 
\end{align*}
The LP in \eqref{eq:LP} with the notation in~\eqref{eq:par2rel} becomes

\begin{align}
\mathsf{d}_{N=2}^{\rm(HD)} 
=  1 +\max
\min\Big\{
  & \lambda_{00} D_1^{(0)}+\lambda_{01} D_1^{(1)}+\lambda_{10} D_1^{(2)}+\lambda_{11} D_1^{(3)}, \nonumber
\\& \lambda_{00} D_2^{(0)}+\lambda_{01} D_2^{(1)}+\lambda_{10} D_2^{(2)}+\lambda_{11} D_2^{(3)}, \nonumber
\\& \lambda_{00} D_3^{(0)}+\lambda_{01} D_3^{(1)}+\lambda_{10} D_3^{(2)}+\lambda_{11} D_3^{(3)}, \nonumber
\\& \lambda_{00} D_4^{(0)}+\lambda_{01} D_4^{(1)}+\lambda_{10} D_4^{(2)}+\lambda_{11} D_4^{(3)}
\Big\},
\label{eq:dofNewNew}
\end{align}
where
\begin{align*}
\begin{array}{ll}
 D_1^{(0)} := \max \left \{ 0, \alpha^\star_{s1},\alpha^\star_{s2} \right \},
 & D_1^{(1)} = D_2^{(0)} := \max \left \{ 0, \alpha^\star_{s1} \right \},
\\D_4^{(3)} := \max \left \{ 0, \alpha^\star_{1d},\alpha^\star_{2d} \right \},
 & D_1^{(2)} = D_3^{(0)} := \max \left \{ 0, \alpha^\star_{s2} \right \},  
\\ D_2^{(1)} :=  \max  \left \{  \alpha^\star_{s1}+\alpha^\star_{2d} +1, \beta_1 \right \},  
 & D_2^{(3)} = D_4^{(1)} := \max \left \{ 0, \alpha_{2d} \right \}, 
\\ D_3^{(2)} := \max \left \{ \alpha^\star_{s2}+\alpha^\star_{1d} +1,\beta_2  \right \},
 & D_3^{(3)} = D_4^{(2)} :=\max \left \{ 0, \alpha^\star_{1d} \right \},
\\ D_1^{(3)} = D_2^{(2)} = D_3^{(1)} = D_4^{(0)} := 0.  
\end{array}
\end{align*}
%\begin{align*}
%\begin{array}{ll}
% D_1^{(0)} := \max \left \{0,\alpha^\star_{s1},\alpha^\star_{s2} \right \},
%& D_1^{(1)} = D_2^{(0)} := \max \left \{0,\alpha^\star_{s1} \right \},
%\\ D_1^{(2)} = D_3^{(0)} := \max \left \{ 0,\alpha^\star_{s2} \right \},  
%& D_1^{(3)} = D_2^{(2)} = D_3^{(1)} = D_4^{(0)} := 0,   
%\\ D_2^{(1)} :=  \max  \left \{  \alpha^\star_{s1}+\alpha^\star_{2d}+1 , \beta_1 \right \},  
%&D_2^{(3)} = D_4^{(1)} := \max \left \{ 0,\alpha^\star_{2d} \right \}, 
%\\ D_3^{(2)} := \max \left \{ \alpha^\star_{s2}+\alpha^\star_{1d}+1 ,\beta_2 \right \},
%&D_3^{(3)} = D_4^{(2)} :=\max \left \{ 0,\alpha^\star_{1d} \right \},
%\\D_4^{(3)} := \max \left \{ 0,\alpha^\star_{1d},\alpha^\star_{2d} \right \}.
%\end{array}
%\end{align*}
Now we have to consider the different cases that may arise:
\begin{itemize}

\item  {\bf {Case 1}}: $\max \left\{ 0,\alpha^\star_{s1},\alpha^\star_{s2}\right\}=0 \Longleftrightarrow \alpha^\star_{s1}\leq0, \alpha^\star_{s2}\leq0$. In this case we have:
\begin{align*}
 D_1^{(0)} =  D_2^{(0)} =  D_3^{(0)}= D_4^{(0)} = 0,
\end{align*}
that is $\lambda_{00}$ is not involved in the optimization; hence, we can set $\lambda_{00}=0$  without loss of optimality.

\item {\bf {Case 2 a}}: $\alpha^\star_{s1}> 0, \alpha^\star_{s2}\leq0$. In this case we have
\begin{align*}
&\mathsf{d} 
= 1  + \max_{(\lambda_{00},\lambda_{01},\lambda_{10},\lambda_{11})\in[0,1]^4 : \lambda_{00}+\lambda_{10}+\lambda_{01}+\lambda_{11} \leq 1} \min  \left\{ \right.
\\&\left. \begin{array}{llll}
\lambda_{00}\alpha^\star_{s1} &+\lambda_{01}\alpha^\star_{s1}                     &+0                                 &+0                                                     \\
0                             &+0                                                 &+\lambda_{10}\max\left\{\alpha^\star_{s2}+\alpha^\star_{1d}+1 ,\beta_2\right\}&+\lambda_{11}\max \left\{0,\alpha^\star_{1d}\right\} \\
0                             &+\lambda_{01}\max\left\{0,\alpha^\star_{2d}\right\}&+\lambda_{10}\max\left\{ 0,\alpha^\star_{1d}\right\}                               &+\lambda_{11}\max \left\{0,\alpha^\star_{1d},\alpha^\star_{2d}\right\} \\
\end{array} \right\}  ,
\end{align*} 
since the second constraint in \eqref{eq:dofNewNew} is always greater than the first one.

Assume $(\lambda_{00},\lambda_{01},\lambda_{10},\lambda_{11})$ is optimal with $\lambda_{00} >0 $; 
the solution $(0,\lambda_{00}+\lambda_{01},\lambda_{10},\lambda_{11})$ gives a higher gDoF (the first and the second equations remain the same, the last one is increased);
we reached a contradiction. Hence the optimal solution must have $\lambda_{00}=0$.

\item {\bf {Case 2 b}}: $\alpha^\star_{s1}\leq0, \alpha^\star_{s2}> 0$.
As Case 2 a above but with the role of the sources swapped. 
Also in this case the optimal solution must have $\lambda_{00}=0$.

\item 
So far we showed that if $\min\left\{ \alpha^\star_{s1},\alpha^\star_{s2}\right\}\leq 0$ then $\lambda_{00}=0$ is optimal.
Due to the symmetry of the problem, by swapping $\alpha^\star_{sj}$ with $\alpha^\star_{jd}$, 
if $\min\left\{ \alpha^\star_{1d},\alpha^\star_{2d}\right\}\leq 0$, then $\lambda_{11}=0$ is optimal.

In oder to prove our claim we must consider one last case:
\begin{align}
\min\left\{ \alpha^\star_{s1},\alpha^\star_{s2}\right\}> 0
\ \text{and} \
\min\left\{ \alpha^\star_{1d},\alpha^\star_{2d}\right\}> 0,
\label{eq:what is left open}
\end{align}
that is when all the links from the source to the relays and all links from the relays to the destination are strictly larger than the direct link.

In order to prove our claim, we must partition the set of parameters in~\eqref{eq:what is left open} into two regimes, say $O_0$ and $O_1$, where in regime $O_0$ we show $\lambda_{00}=0$ is optimal and in regime $O_1$ that $\lambda_{11}=0$ is optimal.
By the symmetry of the problem when swapping $\alpha^\star_{sj}$ with $\alpha^\star_{jd}$, 
the regime $O_0$ must be equal to regime $O_1$ when $\alpha^\star_{sj}$ and $\alpha^\star_{jd}$ are swapped.
Next we show that
\[
O_0=\{\alpha^\star_{s1}\alpha^\star_{s2}\geq\alpha^\star_{1d}\alpha^\star_{2d}\}, \
O_1=\{\alpha^\star_{s1}\alpha^\star_{s2}\leq\alpha^\star_{1d}\alpha^\star_{2d}\}.
\]

\item  {\bf {Case 3}}:
$\alpha^\star_{s1}>0, \alpha^\star_{s2}> 0, \alpha^\star_{1d}>0,\alpha^\star_{2d}>0$. Without loss of generality, we assume $\alpha_{1d}\geq \alpha_{2d}$; in this case we have
\begin{align*}
\begin{array}{ll}
 D_1^{(0)} := \max \left \{\alpha^\star_{s1},\alpha^\star_{s2} \right \},
& D_1^{(1)} = D_2^{(0)} := \alpha^\star_{s1},
\\ D_1^{(2)} = D_3^{(0)} := \alpha^\star_{s2},  
& D_1^{(3)} = D_2^{(2)} = D_3^{(1)} = D_4^{(0)} := 0,   
\\ D_2^{(1)} :=  \max  \left \{  \alpha^\star_{s1}+\alpha^\star_{2d}+1 , \beta_1 \right \},  
&D_2^{(3)} = D_4^{(1)} := \alpha^\star_{2d}, 
\\ D_3^{(2)} := \max \left \{ \alpha^\star_{s2}+\alpha^\star_{1d}+1 ,\beta_2 \right \},
&D_3^{(3)} = D_4^{(2)} = D_4^{(3)} :=\alpha^\star_{1d}.
\end{array}
\end{align*}
Now we aim to find the conditions under which  setting $\lambda_{00}=0$ increases the gDoF compared to a case where $\lambda_{00}>0$.
Finding these conditions is equivalent to solve a system where $\lambda_{00}$ is now split into three parts, that we name $\lambda^\star_{01}$, $\lambda^\star_{10}$  and $\lambda^\star_{11}$. In other words, our aim is to demonstrate that $\left(0,\lambda_{01}+\lambda^\star_{01},\lambda_{10}+\lambda^\star_{10} ,\lambda_{11}+\lambda^\star_{11}\right)$ gives a larger gDoF than
 $\left(\lambda_{00},\lambda_{01},\lambda_{10},\lambda_{11}\right)$ with $\lambda_{00} = \lambda^\star_{01}+\lambda^\star_{10}+\lambda^\star_{11}$.
This is equivalent to solve 
\begin{align*}
&x \max \left\{ \alpha^\star_{s1},\alpha^\star_{s2} \right\} \leq \lambda^\star_{01}\alpha^\star_{s1}+\lambda^\star_{10}\alpha^\star_{s2}
\\& x \alpha^\star_{s2} \leq \lambda^\star_{10}\max\left\{\alpha^\star_{s2}+\alpha^\star_{1d}+1 ,\beta_2\right\}+\lambda^\star_{11}\alpha^\star_{1d}
\\& x \alpha^\star_{s1} \leq \lambda^\star_{01}\max \left\{\alpha^\star_{s1}+\alpha^\star_{2d}+1 ,\beta_1\right\}+\lambda^\star_{11}\alpha^\star_{2d}
\\& 0 \leq \lambda^\star_{01}\alpha^\star_{2d}+\lambda^\star_{10}\alpha^\star_{1d}+\lambda^\star_{11}\alpha^\star_{1d},
\end{align*}
%\label{eq:MDconditions}
%\end{subequations}
where $\lambda^\star_{01}+\lambda^\star_{10}+\lambda^\star_{11}=x$. Now, by substituting $\lambda^\star_{11} = x-\lambda^\star_{01}-\lambda^\star_{10}$, we obtain
\begin{align*}
&x \max \left\{ \alpha^\star_{s1},\alpha^\star_{s2} \right\} \leq \lambda^\star_{01}\alpha^\star_{s1}+\lambda^\star_{10}\alpha^\star_{s2}
\\& x \left (\alpha^\star_{s2}-\alpha^\star_{1d}\right ) \leq \lambda^\star_{10}\max\left\{\alpha^\star_{s2}+1 ,\beta_2-\alpha^\star_{1d}\right\}-\lambda^\star_{01}\alpha^\star_{1d}
\\& x \left (\alpha^\star_{s1}-\alpha^\star_{2d}\right )  \leq \lambda^\star_{01}\max \left\{\alpha^\star_{s1}+1 ,\beta_1-\alpha^\star_{2d}\right\}-\lambda^\star_{10}\alpha^\star_{2d}
\\& \lambda^\star_{01} \leq \frac{x\alpha^\star_{1d}}{\alpha^\star_{1d}-\alpha^\star_{2d}}.
\end{align*}
Notice that, in the last inequality, $\lambda^\star_{01}\leq 1$ holds if there exists a $0\leq x\leq 1$ such that $x \leq \frac{\alpha_{1d}-\alpha_{2d}}{\alpha_{1d}}$ and this is always true since $\frac{\alpha_{1d}-\alpha_{2d}}{\alpha_{1d}} \leq 1$.
Assume equality in the last constraint and substitute the value of $\lambda^\star_{01}$ in the other inequalities. We obtain
\begin{align}
& \lambda^\star_{10} \geq x \frac{\max \left\{ \alpha^\star_{s1},\alpha^\star_{s2} \right\}\alpha^\star_{1d}-\max \left\{ \alpha^\star_{s1},\alpha^\star_{s2} \right\}\alpha^\star_{2d}-\alpha^\star_{s1}\alpha^\star_{1d}}{\alpha^\star_{s2}\alpha^\star_{1d}-\alpha^\star_{s2}\alpha^\star_{2d}}:=\mathsf{A} 
\\& \lambda^\star_{10} \geq x \frac{\alpha^\star_{s2}\alpha^\star_{1d}-\alpha^\star_{s2}\alpha^\star_{2d}+\alpha^\star_{1d}\alpha^\star_{2d}}{\max\left\{\alpha^\star_{s2}+1 ,\beta_2-\alpha^\star_{1d}\right\}\alpha^\star_{1d}-\max\left\{\alpha^\star_{s2}+1 ,\beta_2-\alpha^\star_{1d}\right\}\alpha^\star_{2d}} :=\mathsf{B}
\\& \lambda^\star_{10} \leq x \frac{\max\left\{\alpha^\star_{2d}+1 ,\beta_1-\alpha^\star_{s1}\right\}\alpha^\star_{1d}+\alpha^\star_{s1}{\alpha^\star_{2d}-\alpha^\star_{2d}}^2}{\alpha^\star_{1d}\alpha^\star_{2d}-{\alpha^\star_{2d}}^2}:=\mathsf{C}
\\& \lambda^\star_{01} = \frac{x\alpha^\star_{1d}}{\alpha^\star_{1d}-\alpha^\star_{2d}}.
\end{align}
Thus we should have
\begin{align*}
 \max \left\{\mathsf{A},\mathsf{B} \right\}\leq\lambda^\star_{10}, \quad
 \lambda^\star_{10} \leq \mathsf{C} 
\end{align*}
which is possible if
$ \max \left\{\mathsf{A},\mathsf{B} \right\}\leq \mathsf{C}.$
%In other words, we should find the channel conditions such that $\max \left\{\mathsf{A},\mathsf{B} \right\}\leq \mathsf{C}$ holds.

We notice that $\mathsf{A} \leq \mathsf{C}$ always holds since: (i) if $\max \left\{ \alpha^\star_{s1},\alpha^\star_{s2} \right\}=\alpha^\star_{s1}$, $\mathsf{A}$ is always negative (while $\mathsf{C}$ is always positive); (ii) if $\max \left\{ \alpha^\star_{s1},\alpha^\star_{s2} \right\}=\alpha^\star_{s2}$, then
\begin{align*}
\alpha^\star_{s2} \alpha^\star_{1d} \max \left \{ \alpha^\star_{2d}+1 ,\beta_1-\alpha^\star_{s1} \right \}+\alpha^\star_{s2}\alpha^\star_{s1}\alpha^\star_{2d}&\geq \alpha^\star_{s2} \alpha^\star_{1d} \left ( \alpha^\star_{2d}+1  \right )+\alpha^\star_{s2}\alpha^\star_{s1}\alpha^\star_{2d}
\\&\geq \alpha^\star_{s2} \alpha^\star_{1d} \alpha^\star_{2d} - \alpha^\star_{s1} \alpha^\star_{1d} \alpha^\star_{2d}.
\end{align*}
Thus, our analysis reduces to find the channel conditions such that $\mathsf{B}\leq \mathsf{C}$. 
%{\red why? explain why $\mathsf{A}\leq \mathsf{C}$ always}. 
We must verify for which value of the channel parameters the following inequality holds
\begin{align*}
LHS:=
\max\left\{\alpha^\star_{s2}+1 ,\beta_2-\alpha^\star_{1d}\right\}\left( \max\left\{\alpha^\star_{2d}+1 ,\beta_1-\alpha^\star_{s1}\right\}\alpha^\star_{1d}+\alpha^\star_{s1}{\alpha^\star_{2d}-\alpha^\star_{2d}}^2\right)
\\
\geq
{\alpha^\star_{2d}} \left( \alpha^\star_{s2}\alpha^\star_{1d}-\alpha^\star_{s2}\alpha^\star_{2d}+\alpha^\star_{1d}\alpha^\star_{2d}\right)=:RHS.
\end{align*}
Then the LHS of the inequality above, by using the fact that
\begin{align*}
\max\left\{\alpha^\star_{s2}+1 ,\beta_2-\alpha^\star_{1d}\right\} \geq \alpha^\star_{s2}, \\
\max\left\{\alpha^\star_{2d}+1 ,\beta_1-\alpha^\star_{s1}\right\} \geq \alpha^\star_{2d},
\end{align*}
can be upper bounded as
\begin{align*}
&\max\left\{\alpha^\star_{s2}+1 ,\beta_2-\alpha^\star_{1d}\right\}\left( 
 \max\left\{\alpha^\star_{2d}+1 ,\beta_1-\alpha^\star_{s1}\right\}\alpha^\star_{1d}+\alpha^\star_{s1}{\alpha^\star_{2d}-\alpha^\star_{2d}}^2\right)
\\&\geq 
\alpha^\star_{s2}\left( 
 \alpha^\star_{2d}\alpha^\star_{1d}+\alpha^\star_{s1}{\alpha^\star_{2d}-\alpha^\star_{2d}}^2\right) =: LHS'.
%%
%%\\& \geq \max\left\{\alpha^\star_{s2}+1 ,\beta_2-\alpha^\star_{1d}\right\}\left( \left(\alpha^\star_{2d}+1  \right ) \alpha^\star_{1d}+\alpha^\star_{s1}{\alpha^\star_{2d}-\alpha^\star_{2d}}^2\right)
%%\\& \geq \max\left\{\alpha^\star_{s2}+1 ,\beta_2-\alpha^\star_{1d}\right\}\left(\alpha^\star_{2d} \left( \alpha^\star_{1d}-\alpha^\star_{2d} \right)+\alpha^\star_{s1}\alpha^\star_{2d}\right)
%%\\& \geq \left (\alpha^\star_{s2}+1  \right )\left(\alpha^\star_{2d} \left( \alpha^\star_{1d}-\alpha^\star_{2d} \right)+\alpha^\star_{s1}\alpha^\star_{2d}\right)
%%\\& \geq \alpha^\star_{s2}\left(\alpha^\star_{2d} \left( \alpha^\star_{1d}-\alpha^\star_{2d} \right)+\alpha^\star_{s1}\alpha^\star_{2d}\right)
%\\& \geq \alpha^\star_{2d}\left(\alpha^\star_{s2}\alpha^\star_{1d}-\alpha^\star_{s2}\alpha^\star_{2d}+\alpha^\star_{1d}\alpha^\star_{2d}\right)\Longleftrightarrow \alpha^\star_{s1}\alpha^\star_{s2}\geq\alpha^\star_{1d}\alpha^\star_{2d}.
\end{align*}
If $LHS' \geq RHS$ then also $LHS \geq RHS$; therefore ,
\begin{align*}
&\alpha^\star_{s2}\left( 
 \alpha^\star_{2d}\alpha^\star_{1d}+\alpha^\star_{s1}{\alpha^\star_{2d}-\alpha^\star_{2d}}^2\right)
 \geq \alpha^\star_{2d}\left(\alpha^\star_{s2}\alpha^\star_{1d}-\alpha^\star_{s2}\alpha^\star_{2d}+\alpha^\star_{1d}\alpha^\star_{2d}\right)
\\&\Longleftrightarrow \alpha^\star_{s1}\alpha^\star_{s2}\geq\alpha^\star_{1d}\alpha^\star_{2d}.
\end{align*}
%Thus, $\lambda_{00}=0 \Longleftrightarrow \alpha^\star_{s1}\alpha^\star_{s2}\geq\alpha^\star_{1d}\alpha^\star_{2d}$.
Thereby, we can draw the following conclusion: assume $(\lambda_{00},\lambda_{01},\lambda_{10},\lambda_{11})$ is optimal with $\lambda_{00} >0 $; 
as demonstrated above, the solution $(0,\lambda_{01}+\lambda^\star_{01},\lambda_{10}+\lambda^\star_{10},\lambda_{11}+\lambda^\star_{11})$ gives a higher gDoF;
we reached a contradiction. Hence the optimal solution must have $\lambda_{00}=0$.
%{\red the proof is not over: you still need to show that the conditions in~\eqref{eq:MDconditions} allow you to find a solution that is better than the one with lambda00=0, i.e., as I did before ``Assume $(\lambda_{00},\lambda_{01},\lambda_{10},\lambda_{11})$ is optimal with $\lambda_{00} >0 $; 
%the solution $(0,\lambda_{00}+\lambda_{10},\lambda_{01},\lambda_{11})$ gives a higher gDoF (first and second eq. the same, last one increased);
%we reached a contradiction. Hence the optimal solution must have $\lambda_{00}=0$. ''
%}

By the same reasoning,  $\lambda_{11}=0 \Longleftrightarrow \alpha^\star_{s1}\alpha^\star_{s2}\leq\alpha^\star_{1d}\alpha^\star_{2d}$.

\end{itemize}

\bibliographystyle{IEEEbib}
\bibliography{IccBib}

%\begin{figure*}
%\centering
%\includegraphics[width=0.7\textwidth]{BestVSBoth.pdf}
%\caption{Example of a network when the condition in~\eqref{eq:were we do better than relay selection case 2a} holds with: $\alpha_{s1}=1.3$, $\alpha_{s2}=0.6$, $\alpha_{1d}=0.3$, $\alpha_{2d}=1.7$, $\beta_1=1.4$ and varying $\beta_2$. {\red remove legend; use black color with 2 different line styles; add description here..}}
%\label{fig:BestVSBoth}
%\end{figure*}

\end{document}